\documentclass[iop]{emulateapj}
\shortauthors{Sekanina}
\shorttitle{Aristotle's Comet}
\slugcomment{Version \today }

\begin{document}
\title{New Insights into the Nature and Orbital Motion of Aristotle's Comet
 in 372 BC\\[-1.55cm]}
\author{Zdenek Sekanina}
\affil{La Canada Flintridge, California 91011, U.S.A.; {\sl ZdenSek@gmail.com}}
\begin{abstract} 
Extending the investigation of the presumed primordial comet as part of
continuing work~on~a~new model of the Kreutz sungrazer system, I confront a
previously derived set of orbital elements with~Aristotle's remarks in his
{\it Meteorologica\/} to test their compatibility and determine the comet's
perihelion time.  The two translations of the treatise into English that I am
familiar with differ at one point substantially from each other. Unambiguously,
the year and season of the comet's~\mbox{appearance}~was~early 372~BC (or
$-$371).  From Aristotle's constraint on the comet's setting relative to
sunset, I infer that the~probable date of perihelion passage was January~20, a
date also consistent~with~the~vague~remark on frosty weather.  On the day that
Aristotle claims the comet was not seen, its~head~may~have been hidden behind
the Sun's disk or in contact with it.  The observation that the ``{\it comet
receded as~far~as Orion's belt, where it dissolved\,\/}'' is being satisfied
by the tested orbit if the perihelion was reached between January~20 and
February~10.  Aristotle's third statement, which describes the tail~as~a~streak
60$^\circ$ in length, suggests a plasma feature stretching in space
over 0.8~AU.  The dust tail was developing more gradually and it was all
that could be seen from the comet when it was approaching Orion's belt in
early April.  The comet was seen over a period of more than 10~weeks.  The
results of this study strengthen the notion that Aristotle's comet indeed
was the gigantic progenitor~of~Kreutz~\mbox{sungrazers}.
\end{abstract}

\section{Introduction} 
Aristotle's comet appears to be the earliest among the comet
celebrities, having blazed the sky about 130~years before the
first known appearance of Halley's comet.\footnote{There are
unconfirmed records of Halley's comet from 467~BC, nearly a
century before Aristotle's comet (e.g., Ho 1962 and references
therein), but the first confirmed return was in 240~BC.}  Like
Edmond Halley, Aristotle did not discover ``his'' comet, but
unlike Halley, he did not compute~its~\mbox{orbit}.  Aristotle
was, as a boy of 11 or 12, merely one of many Greeks and other
peoples living at the time of the comet's appearance who witnessed
its display.  His advantage over millions of other spectators
was that he grew to become the foremost philosopher of his time,
who decades later wrote --- among numerous other works --- his
famous {\it Meteorologica\/}.
In this treatise he put down a few lines on the comet as part
of a discourse on the nature of these objects.

Although Aristotle was not the only highly educated Greek who
saw the comet, he provided information that proved truly useful,
as shown in this paper.  That cannot be said about Ephorus of
Cyme, a historian who described his seeing the comet to split
up into two planets.  While fragmentation of this comet at or near
perihelion could by no means be ruled out (it was in fact likely,
as discussed in the following), no human being could possibly
see such as event taking place with the naked eyes.  Nucleus
fragmentation is a process whose {\it consequences\/} can only
be followed telescopically and with much delay.  To witness
it in its workings, one needs an imager on board a spacecraft
positioned in the immediate proximity of the comet's nucleus at
the time it is fragmenting.  Some outreach texts on Aristotle's
comet remark that Callisthenes of Olynthus, another historian,
commented on it as well, but since he was born more than
10~years after the comet's appearance, his contribution could
only have been hearsay.

Additional remarks on Aristotle's comet were offered by
historians who lived centuries later, and I see no input
of much assistance to this study from any of them.  One of these
was Diodorus Siculus, whose account I will comment on in Section~2.

One peculiar point about Aristotle's comet was that there has
apparently been no record of it from China.  This may have to do
with Qin Shi Huang, China's first emperor known for the Terracotta
Army collection, who is said to have ordered to burn most books in
213~BC.  Modern comet catalogues do indeed show fewer Chinese
sightings of comets before the late 3rd century BC compared to
sightings elsewhere; for example, Hasegawa's (1980) catalogue
suggests that 57~percent of recorded comets came from China
between 1~BC and 200~BC, but only 41~percent between 301~BC and
500~BC.

In any case, Aristotle's narrative is the only firsthand account of
the comet's appearance that I will rely upon in this investigation.
In the following the aim is to examine his remarks in detail.

\section{The Appearance of Aristotle's Comet}
The words that Aristotle used in describing the comet are of key
importance.  Yet, one immediately confronts a problem:\ it is not
Aristotle's words that one works with, but a text translated from
Greek into English.

To start with, I examined the section on the comet~in Kronk's (1999)
{\it Cometography\/} as well as the pertinent paragraph in Seargent's
(2009) book on {\it The Greatest Comets in History\/}.  There are
numerous translations of Aristotle's account (Barrett 1978 offers
another~one), but
comparison shows that the versions presented by Kronk and Seargent
differ by exactly one word.  For the reader's benefit,
here is the relevant text from Seargent's book:

\parbox{8cm}{\it \small \vspace{0.145cm}The great comet, which appeared
about the time of the earthquake in Achaea and the tidal wave, rose
in the west.\,\,\ldots The great comet \ldots appeared during winter
in clear frosty weather in the west, in the archonship of Asteius:\
on the first night it was not visible as it set before the sun did,
but it was visible on the second, being the least distance behind
the sun that would allow it to be seen, and setting immediately.
Its light stretched across a third of the sky in a great band, as
it were, and so was called a path.  It rose as high as Orion belt,
and there disappeared.\\}

In the translation used by Kronk the last word was {\it dispersed\/}
rather than {\it disappeared\/}, the only difference between the
two versions.  On the other hand, I find out that Webster (2004)
has translated this segment of Aristotle's text as follows: \\

\parbox{8cm}
{\it \small The great comet which appeared at the time of the
earthquake in Achaea and the tidal wave rose due west;\,\,\ldots
the great comet appeared to the west in winter in frosty weather
when the sky was clear, in the archonship of Asteius.  On the
first day it set before the sun and was then not seen.  On the
next day it was seen, being ever so little behind the sun and
immediately setting.  But its light extended over a third part
of the sky like a leap, so that people called it a `path'.  This
comet receded as far as Orion's belt and there dissolved.\\}

Although far from identical, these two translations do convey essentially
the same information, with one major exception.  At the beginning
of the last sentence the first version uses the words {\it it rose\/}
while Webster's version says {\it this comet receded\/}.  This
difference is significant:\ since the previous sentence deals
with the comet's tail, one expects that {\it it rose\/} alludes to
the tail as well.  However, Webster's translation leaves no doubt
that in the last sentence Aristotle talks about the comet's motion,
not about its tail orientation.  I should mention that in reference
to the translation he used, Seargent expressed his misgivings
about exactly this point.  He noted that the last sentence was
ambiguous and wondered whether the statement was meant to refer
to the tail or the comet as a whole.

\section{Year of Arrival of Aristotle's Comet}
In the previous sections I deliberately avoided a direct
reference to the year of Aristotle's comet, the subject~of this
section.  Having addressed this problem in the past (Sekanina
2022a), I concluded that the comet undoubtedly appeared in early
372~BC (or $-$371).  I now summarize my main points, including
the context of why~the~incorrect year 371~BC has kept showing up
in the literature.  Any controversy about the year is prevented
by carefully reading Aristotle's words.

In particular, his statement that the comet appeared when Asteius
was archon settles this problem unequivocally.  It confirms that
Aristotle used the Athenian calendar, in which every year started in
midsummer, after the solstice, with the lunar month of Hekatombaion.
The Athenian chief magistrate, called {\it eponymous archon\/}, was
in office for a period of one year and his name correlated with
one year in the four-year long Olympiad cycle.  Asteius' archonship
coincided with the 4th year of the 101st Olympiad, which in our
calendar lasted from July 373~BC to June 372~BC.
 
It appears that one source responsible~for~the~incorrect year,
371~BC, was a secondhand account~of~the~comet by Diodorus Siculus,
written about three centuries~later.  Unfortunately, in glaring
contradiction to Aristotle's eyewitness description, Diodorus
re\-marked in his {\it Biblioteca Historica\/}~15.50 that
`a great blaz\-ing torch \ldots was seen in the heavens \ldots
{\it when Alcisthenes was archon at Athens\/}.'  Alcisthenes
followed Asteius as archon in the first year of the 102nd
Olympiad, equivalent to the period from July 372~BC to June 371~BC.
We will never know what made Diodorus move the time of the comet's
arrival forward by one year, but one possibility is that it was his
obsession with portents.  Diodorus believed that the comet was a
`divine portent' that foretold the defeat of Spartans by Boeotians
in the Battle of Leuctra in July 371~BC.  Adjusting the time of
sighting deliberately as done, the comet's appearance was by that
amount of time closer to the time of the battle, thereby becoming
a much more `relevant' portent.  Pure fabrication \ldots

There was another event whose timing independently confirms that
Aristotle's comet could not have appeared in 371~BC --- the Achaean
earthquake causing destruction of the city-states of Helike, which
was submerged by a tsunami, and Boura, which collapsed and caved
in.  As the excerpts from his treatise show, Aristotle remarked
that the comet had appeared at or about the time of this
earthquake.  Historical accounts by others (Herakleides of Pontos,
Theophrastos, Eratosthenes, Poseidippos of Pella, Aristophanes
Byzantios, etc.) suggest quite consistently that the cataclysmic
earthquake took place in winter of 373/372~BC (e.g., Kolia 2011,
Katsonopoulou \& Koukouvelas 2022), so that the two events appear
to have happened just weeks apart.{\vspace{-0.1cm}}

\section{Orbit of Aristotle's Comet}
Although Aristotle's account of the comet's motion is
clearly short of what one would expect as minimum necessary
information to get an idea about the orbital properties (except
that the perihelion distance was very small and the motion
retrograde), Pingr\'e (1783)~\mbox{managed} to come up with further
orbital constraints, however crude, at a time when the existence
of the sungrazer group was yet to be recognized.

In the notes accompanying his catalogue of cometary orbits,
Cooper (1852) extensively discussed and elaborated on Pingr\'e's
comments regarding Aristotle's narrative on the comet.  Both
Pingr\'e and Cooper appear to have been particularly intrigued
by the claim that the comet's head was not seen on the first day
as it set before the Sun, an issue I will examine in Section~5.
Pingr\'e apparently had a problem with the tail in front of
the head, which he (mistakenly) deemed `impossible.'~Cooper
then contemplated the position of the descending node and
the comet's disappearance as far as the `girdle' of Orion.
To quantitatively express Pingr\'e's conjectures on the
comet's motion, Cooper introduced in his catalogue 60$^\circ$
wide intervals of uncertainty in the longitude of perihelion
and longitude of the ascending node, which in combination
implied a 120$^\circ$ wide interval of uncertainty in the
argument of perihelion.\footnote{The longitude of perihelion,
$\pi$, an orbital element used in early computations but now
replaced by the argument of perihelion, $\omega$, is not to
be confused with the ecliptic longitude of perihelion,
$L_\pi$, one of two quantities defining the direction of
the line of apsides.}

\begin{table}[t] 
\vspace{0.19cm}
\hspace{-0.21cm}
\centerline{
\scalebox{1}{
\includegraphics{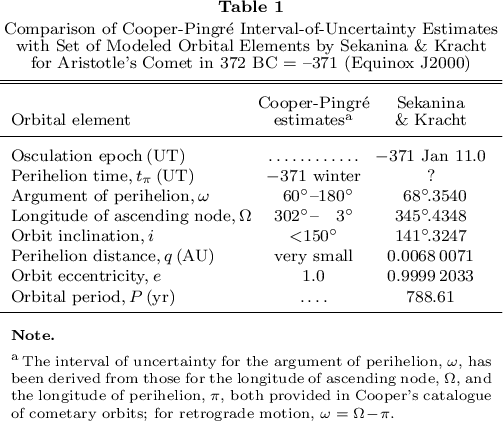}}}
\vspace{0.35cm}
\end{table}

Pingr\'e's and Cooper's results were, to my knowledge, essentially
all that had been available on the subject before our recent
effort (Sekanina \& Kracht 2022), aimed at supporting the new
contact-binary model for the Kreutz system (Sekanina 2021)
by integrating the motions~of~the major sungrazing comets over a
period of two millennia.  The high-precision set of predicted
orbital elements for Aristotle's comet was a byproduct of
this project, as we extended the computations by another two
millennia.

Remarkably, the Cooper-Pingr\'e benevolent allowances for the
uncertainties in the orbital motion of Aristotle's comet happen
to cover the rigorously derived set of elements, compatible with
the contact-binary model and presented in Table~1.  The set of
five elements~in~the rightmost column of the table --- $\omega$,
$\Omega$, $i$, $q$, and $e$ --- is considered in the following
as firmly established and used throughout this paper.  The only
element that still proves somewhat elusive is the perihelion
time.  Preliminary constraints, based on comparisons with recent
members of the Kreutz system suggested that the comet reached
perihelion in February or early March of 372~BC and that the
uncertainty of this result was probably less than $\pm$1~month
(Sekanina 2022a).  Yet, more tests have been desirable and a
refinement of the perihelion time of Aristotle's comet is a
subject of this investigation.

The tabulated elements indicate that the Earth was crossing the
comet's orbital plane on February~5.6~UT, 372~BC.  The winter
solstice was on December~25.4 UT, 373~BC and the spring equinox
on March~25.8, 372~BC.  The new moon occurred on January~16.3,
February~15.1, and March 16.6~UT, 372~BC.  To avoid bias, I
consider five dates, with a 20-day step, that cover entire winter,
the period of time between the solstice and equinox, as the starting
hypotheses for the osculating perihelion time (at an epoch of
January~11.0~UT, 372~BC; see Table~1):\ January~1.0, January~21.0,
February~10.0, March~2.0, and March~22.0~UT (JD\,1585550.5--1585630.5).

Given that Aristotle wrote his treatise some forty or so
years after the comet's appearance, his comments are in some
respects remarkably informative, although somewhat conjectural.
These doubts concern especially his claim that on the first
of two nights the comet was not seen because it had set prior
to the Sun, the issue that Pingr\'e seemed confused about.
Of course, one does readily recognize the speculative nature
of a statement on the position of something that has not
been seen.  But even if, perhaps improbably, the sentence was
meant to imply that on the first night the comet's tail was
seen~{\it \mbox{without} the head\/}, the comet's early setting
may not necessarily have been the cause.  Both deteriorated
atmospheric transparency and twilight could have made it essentially
impossible for the comet's head to be seen extremely low near the
horizon.  One also could question whether {\it the next day\/} was
indeed meant 24~hours later, or the day of the next opportunity
to see the comet, if there were one or more cloudy nights in between.
The chance that one day no comet at all was seen and 24~hours later
its tail extended over a third of the sky should probably be ruled
out.  Instead, a variety of scenarios should be examined from
two angles:\ (i)~as a function of projection conditions that are
dependent on the perihelion time; and (ii)~as an issue of brightness
conditions requiring assumptions on the comet's light curve and
the limiting naked-eye magnitude.  Solutions should result in
some constraint on the perihelion time of Aristotle's comet.

The second constraint is provided by Aristotle's claim that
the comet receded to, and disappeared at, Orion's belt.  At
first sight, this is a surprising statement, given that the
Kreutz sungrazers are known to arrive from and leave for, deep
southern regions of the sky.  It is of course also true that
the ecliptic reaches its peak northern declination at right
ascension of 6$^{\rm h}$, near Orion.\footnote{In Aristotle's
times, the right ascensions of the Orion belt stars were slightly
less than 5$^{\rm h}$.} In any case, the position at Orion's
belt is a robust test of Aristotle's comet as a Kreutz sungrazer.
And, last but not least, the comet's recorded disappearance at
Orion's belt implies a sufficiently large solar elongation at
the time, which needs to be supported by the ephemeris dependent
on the perihelion time.

The third constraint follows from Aristotle's remark that the
tail was 60$^\circ$ long on the `second' night.  While one expects
that this length is likely to be accounted for by the plasma tail,
the words like {\it band\/} or {\it path\/} suggest a certain
sustained width of the feature, so that some limited contribution
from dust is not ruled out.  Another issue is that of the tail's
potential curvature; Aristotle mentioned none.  This circumstance
may suggest that the tail reached its maximum length when the
Earth was close to the comet's orbital plane, in early February.
However, the tail's visibility was also affected by its
orientation in the sky, especially by the deviation of its
direction from the horizon line.  Each of Aristotle's three
constraints is next examined separately.

\section{Aristotle's Comet at Close Proximity to Sun}
Investigation of the first constraint requires the knowledge
of the projected path of Aristotle's comet at close proximity
to the Sun's disk, with the perihelion time as a parameter.  For
each of the five selected scenarios (Section~4), the comet's
predicted near-perihelion arc of its path, as it would have
been seen by a hypothetical Greek observer --- a resident of
Athens (longitude 23$^\circ\!$.63\,E, lati\-tude +38$^\circ\!$.00\,N)
--- is plotted in Figure~1, with a detailed description in the
caption.

\begin{figure*}[t] 
\vspace{0.18cm}
\hspace{-0.22cm}
\centerline{
\scalebox{0.85}{
\includegraphics{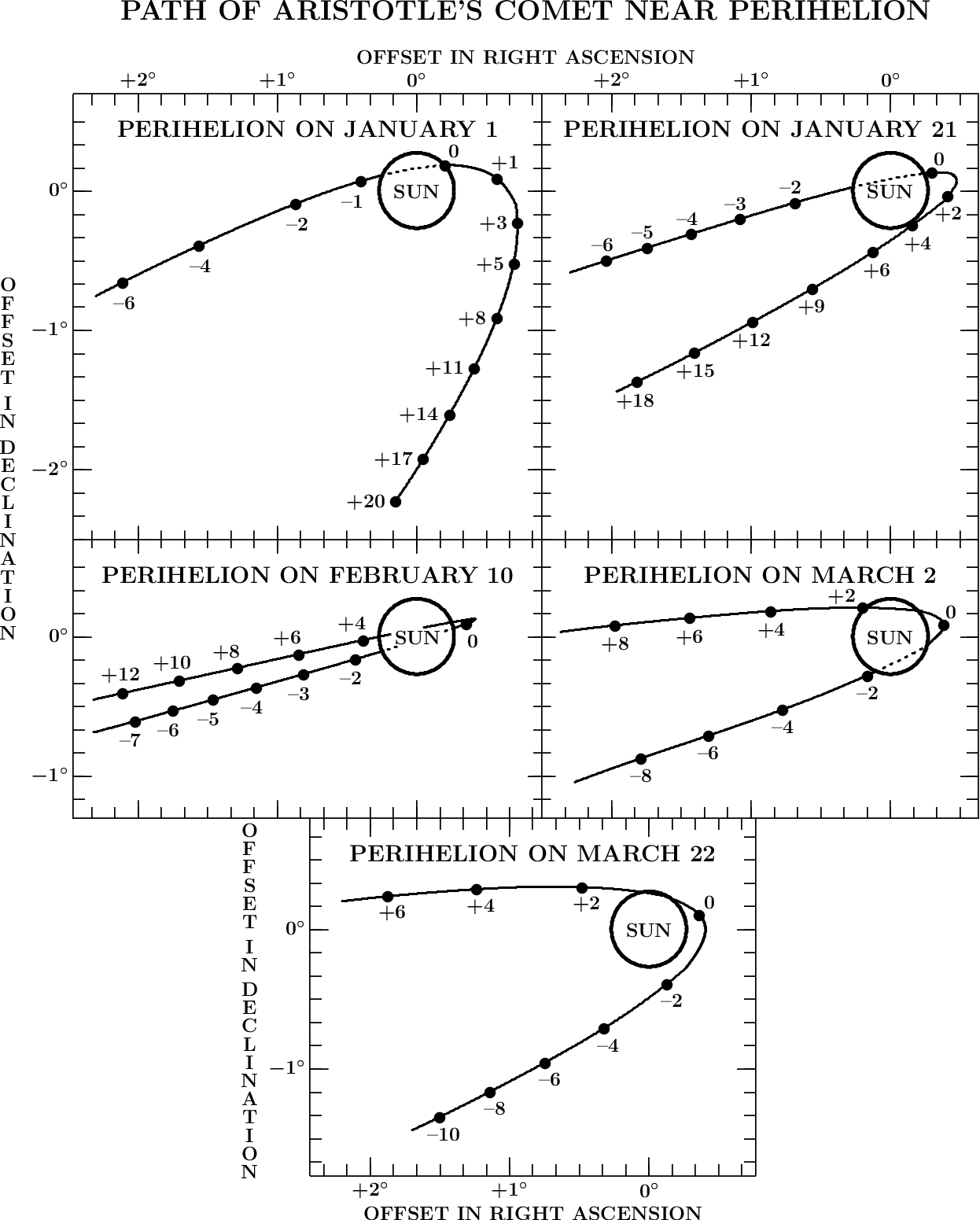}}}
\vspace{0cm}
\caption{Predicted path of Aristotle's comet near perihelion               
relative to the Sun's disk, with the perihelion time 
assumed between January~1 and March 22, 372~BC at a 20-day step.
The solid dots are the comet's offset positions at times, measured
in hours from perihelion (negative = before, positive = after),
in the equatorial coordinate system (equinox J2000) and predicted
for Athens, Greece (longitude 23$^\circ\!$.63\,E, latitude
+38$^\circ\!$.00\,N).  The dotted parts of the path show the
comet behind the Sun's disk, whose size is drawn to scale.}
\end{figure*}

The paths show a wide range of shapes, whose trend with the
perihelion time is in line with the Earth's transit of the
comet's orbital plane on February~5 (Section~4).  In particular,
the comet's projected motion is clockwise in the first two
plots, when perihelion occurred before February~5, but
counterclockwise in the three remaining plots, as the Earth
moved from its positions below the plane to the positions
above it.  The path shape was changing accordingly; it would
project as a straight line in a position angle of 106$^\circ$
(or 286$^\circ$) on February~5.

\subsection{Problem of the Comet's Early Setting}
A major feature of Figure 1 is the comet's consistently
approaching the Sun from the east and receding from~it also
eastward.  As a result, the comet typically set later than
the Sun.  Only in close proximity of perihelion, over a
period of up to several hours in January but merely about
an hour or so in March, was the comet projecting to the
west of the Sun.  Although one should not make any
far-reaching conclusions from the restricted two-dimensional
view that the figure offers, it seems that if the comet
did set, as Aristotle speculated, earlier than the Sun, this
could in all probability have happened only in January and
on the day of perihelion --- which is identical with his
``first day''. 

To investigate more rigorously possible scenarios for the
Athenian observer, I ascertain the time of sunset on the
day of perihelion; determine, in the equatorial coordinate
system for the equinox of the date, the position angle of
the zenithal direction measured from the Sun~at the time;
and equate zenithal distances that exceed 90$^\circ$ with
positions of the comet when setting earlier~than~the Sun.
In the spherical triangle Sun-zenith-celestial pole, the arc
zenith-Sun equals 90$^\circ$ at sunset;\footnote{Atmospheric
refraction, which affects all objects at the same elevation
(or zenithal distance) equally, is ignored.} the arc
Sun-celestial pole equals \mbox{90$^\circ \!\!-\! \delta_\odot$},
where $\delta_\odot$ is the Sun's apparent declination at sunset;
and the arc celestial pole-zenith equals \mbox{90$^\circ \!\!-\!
\phi$}, where $\phi$ is the geographic latitude of the observing
site.  The angles in the spherical triangle opposite the arcs
are, respectively, the sunset's hour angle, $\theta_\odot$;
\mbox{180$^\circ \!\!-\! A_\odot$}, where $A_\odot$ is the
Sun's azimuth at the time of its setting, measured clockwise
from the south; and the position angle of zenith, $P_{\rm z}$,
measured at the Sun --- the quantity that I am looking for.
It is given by the relations
\begin{eqnarray}
\sin P_{\rm z} & = & \frac{\sin A_\odot \cos \phi}{\cos
 \delta_\odot}, \nonumber \\
\cos P_{\rm z} & = & \cos \theta_\odot \cos A_\odot
 + \sin \theta_\odot \sin A_\odot \sin \phi.
\end{eqnarray}
\begin{table}[t] 
\vspace{0.15cm}
\hspace{-0.2cm}
\centerline{
\scalebox{0.997}{
\includegraphics{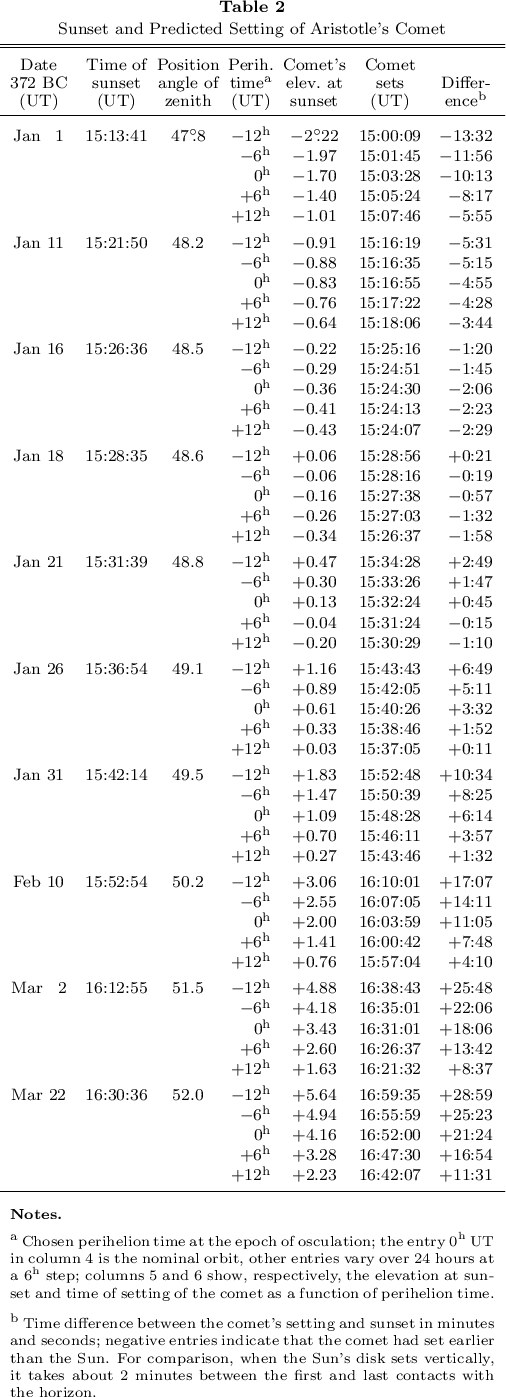}}}
\vspace{-0.2cm}
\end{table}

\begin{table*} 
\vspace{0.15cm}
\hspace{-0.2cm}
\centerline{
\scalebox{1.015}{
\includegraphics{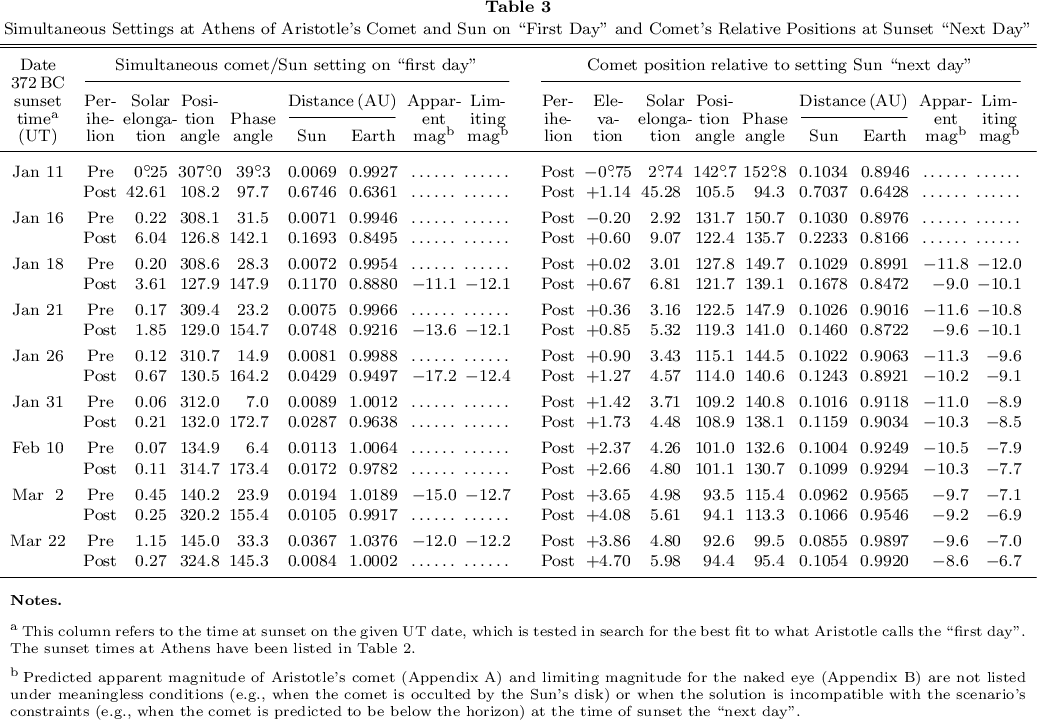}}}
\vspace{0.68cm}
\end{table*}
For each of the five selected perihelion times and five
additional intervening days, Table~2 lists the UT time of
sunset at Athens and the position angle of zenith measured
at the Sun in columns~2 and 3, respectively.  Columns 4--7
show the effect of 6$^{\rm h}$ shifts in the perihelion time
on the comet's elevation at sunset and on the time of the
comet's setting, the quantities that are relevant to
Aristotle's narrative.  The results are very illuminating.
As hinted from my discussion of the zenithal position angle,
the scenarios with the perihelion time in most of January
display a strong propensity for the comet's setting earlier
than the Sun.  This is particularly true about the January~1
scenarios, where I find the comet was setting by as much
as nearly 15~minutes earlier and located up to more than
2$^\circ$ below the horizon at sunset.  The effect is seen
only marginally in most scenarios with the perihelion times
in the second half of January and it disappears completely
in scenarios with the perihelion time at the end of January
or later.  The implication is that if Aristotle's claim on
the comet's early setting on the first day is valid, the
{\it comet reached perihelion not later than the second
half of January\/}.

Besides this rather obvious conclusion, Table~2 displays
a few interesting features to be exploited in the
following.  Five more dates have been added in order
to facilitate tracking the trends in the elevation
of the comet at sunset (column~5) and the difference
between the times of setting of the comet and the Sun
(column~7), which is strongly correlated with the
elevation.  The table shows that for scenarios with
perihelion on January~1 the comet is setting much
earlier than the Sun, but the difference is decreasing
as the perihelion time is advancing from $-$12$^{\rm h}$
to +12$^{\rm h}$.  For January~21, the next of the five
standard dates, the trend changes dramatically.  At
$-$12$^{\rm h}$, the comet is setting nearly 3~minutes
{\it later\/} than the Sun, but the difference changes
its sign before +6$^{\rm h}$ and at +12$^{\rm h}$ the
comet is setting more than 1~minute earlier than the
Sun.  This trend remains unchanged except that the lag
of the comet's setting behind sunset continues to grow.
The three additional entries between January~1 and
January~21 show that the trend is caused by the
systematically changing slope of the curve of elevation
(and time of setting relative to sunset) of the comet
as a function of the date.  The change of sign from
positive to negative is seen to take place between
January~11 and 16, when the comet is setting earlier
than the Sun.  The two additional entries between
January~21 and February~10 confirm that the transition
from the comet-setting-first scenarios to the
Sun-setting-first scenarios is smooth.

\subsection{Simultaneous Settings of the Sun
and the Comet.  Light Curve and Limiting
Naked-Eye Magnitude.}
Given the speculative nature of Aristotle's claim that
the comet set before the Sun on the first day, I will
adopt a more cautious approach and assume that Aristotle
was correct ``half way'':\ the comet set neither before
nor after the Sun, but {\it simultaneously\/} with the
Sun.  And I will link this condition with Aristotle's
following statement, which essentially claimed that next
day the comet set just very shortly after the Sun.

When I searched an interval of perihelion times much wider
than 24~hours, I ascertained that depending on the comet's
position in the orbit (i.e., perihelion time), it could
set simultaneously with the Sun every day twice:\ once before,
and once after, it reached perihelion.  Some key data
for either scenario are provided in Table~3, the individual
columns listing for the time of simultaneous setting on the
``first day'':\ the comet's solar elongation and position
angle, the phase angle, the heliocentric and geocentric
distances, the predicted apparent magnitude, and the
limiting magnitude for the naked eye are, respectively,
on the left-hand side; the data for the comet's position
with respect to the Sun ``next day'':\ the elevation,
as seen from Athens at the time of sunset, plus the same
data as for the ``first day'' are on the right-hand side.
The computation of the comet's apparent magnidude from a
formula for the predicted light curve is described in
Appendix~A, the application of the limiting magnitude
for the naked eye in Appendix~B.

I now carefully examine the positional data in Table~3 and
judge their compatibility with the two conditions that
follow from Aristotle's narrative when taken literally:\
(a)~the comet was not seen on the ``first day'' and (b)~it
set immediately after the Sun on the ``next day''.

{\it January~1\/} is left out because it displays the same
problems as January~11, even more so.

{\it January~11\/}:\ The Preperihelion scenario shows the comet
occulted by the Sun's disk on the ``first day'' and thus
invisible.  However, the ``next day'' the comet~is~predicted
to have been below the horizon, contradicting condition~(b).
The Post-perihelion scenario places the comet more than
40$^\circ$ from the Sun when setting, likewise violating
condition~(b).  This solution is plainly unacceptable.

{\it January~16--18\/}:\ The Preperihelion scenarios suffer from
similar problems as that of January~11.  However, the latter date
has features of a marginally plausible solution.  On the 19th
(``next~day'' of this solution), the Sun and the comet set
essentially simultaneously and the comet's predicted apparent
brightness is near the visibility threshold for the naked eye.
The Post-perihelion scenarios still include extremely large
separations of 7$^\circ$ to 9$^\circ$ from the Sun.  With the
exception of the January~18 preperihelion scenario, which is
close to borderline, the solutions are unacceptable.

{\it January 21--26\/}:\ The Preperihelion scenarios appear
to be in fair agreement with both conditions:\ the comet
was hidden behind the Sun (and therefore invisible) on the
``first day''; and at elevations between 0$^\circ\!$.36 and
0$^\circ\!$.90 and of apparent magnitude of perhaps $-$11
to $-$12 (brighter than the visibility limit) at sunset the
``next day''.  Depending on the interpretation of Aristotle's
remark on the comet's {\it immediately setting\/}, the earlier
or later part of this period is slightly preferable to the
other.  This issue is discussed in greater detail below,
taking into account that for the Sun's disk to set it takes
about three minutes between the first and last contacts with
the horizon.  These two Preperihelion scenarios represent
good starting points to further search for a solution.  The
Post-perihelion scenarios fail because they make the comet
visible on the ``first day'', thereby violating condition~(a).
In addition, the January~21 Post-perihelion scenario also
contradicts condition~(b), making the comet too faint to see
on the ``next day''.

{\it January 31 and later\/}:\ Both Preperihelion and Post-perihelion
scenarios are all incompatible with condition~(b), because
the elevations of about 1$^ \circ\!$.5 or larger the ``next
day'' imply the comet's setting more than 6~minutes after
sunset.  The March~2 Preperihelion scenario also
defies condition~(a).

A tentative conclusion based on conditions (a) and (b) is that
the {\it comet reached perihelion between January~18 and 27\/}.
Aristotle's ``first day'' and ``next day'' appear to refer,
respectively, to times within the last 24~hours before
perihelion and the first 24~hours after perihelion.

\subsection{Possible Date of Aristotle's Initial Sighting\\of the
 Great Comet}
Two kinds of uncertainty are involved in the meaning of
Aristotle's remarks on the comet ``being ever~so~\mbox{little}
behind the sun'' and ``immediately setting.''~It~is~not~only
that one needs to adopt some numbers to express these vague
terms, but also at issue are the finite dimensions of the Sun's
disk in the sky, so that sunset is not an instantaneous event.
I proceed on the assumptions that ``setting immediately'' can be
equated with {\it setting 15 to 45~seconds later\/} and that, with
further reference to the comet's setting,``behind the Sun'' means
after the {\it last contact\/} of the setting Sun with the horizon.
Since in Tables~2 and 3 sunset is referred to the center of the
Sun's disk at horizon (or zero elevation), whose radius equaled
0$^\circ\!$.27 (to within a few arcsec) on the relevant dates,
the time of last contact is equivalent to the time when the center
of the Sun's disk reached an elevation (or, rather, depression) of
$-$0$^\circ\!$.27.

The task is now to test the dates in the second half of January
for complying with an orbital position of the comet such that
the ``next day'' it set 15 to 45~seconds after the last contact
of the Sun with the horizon, while the ``first day'' it set as
early as possible, preferably before the first contact of the
Sun's disk with the horizon.  The results, summarized in Table~4,
show that only on one day, {\it January~20\/}, did the comet
satisfy this particular condition for both scenarios (referred
to in the table as case\,``15'' and case\,``45'', respectively)
and that the time differences are extremely minor, at most
15~seconds.  Nonetheless, Aristotle appears to have been
technically correct when he claimed that the comet set before
the Sun on the ``first day''.  Table~4 suggests that this day
was quite possibly {\it January~20\/}.  If the reference time
is taken to be the contact of the Sun's center (rather than
the first contact) with the horizon, the solution is not
unequivocal.  Yet, it is limited to a period of only a few
days ending with January~20.

\begin{table}[b] 
\vspace{0.5cm}
\hspace{-0.22cm}
\centerline{
\scalebox{1}{
\includegraphics{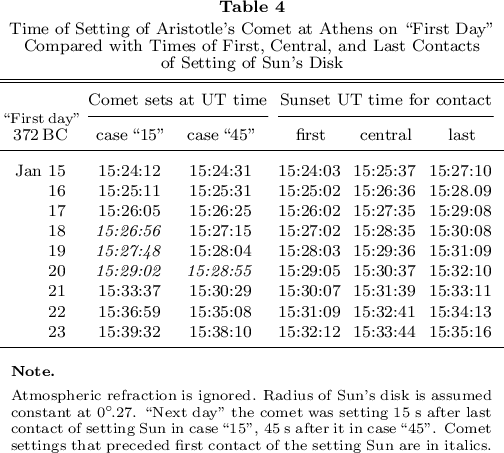}}}
\vspace{-0.08cm}
\end{table}
\begin{table*}[t] 
\vspace{0.15cm}
\hspace{-0.2cm}
\centerline{
\scalebox{0.975}{
\includegraphics{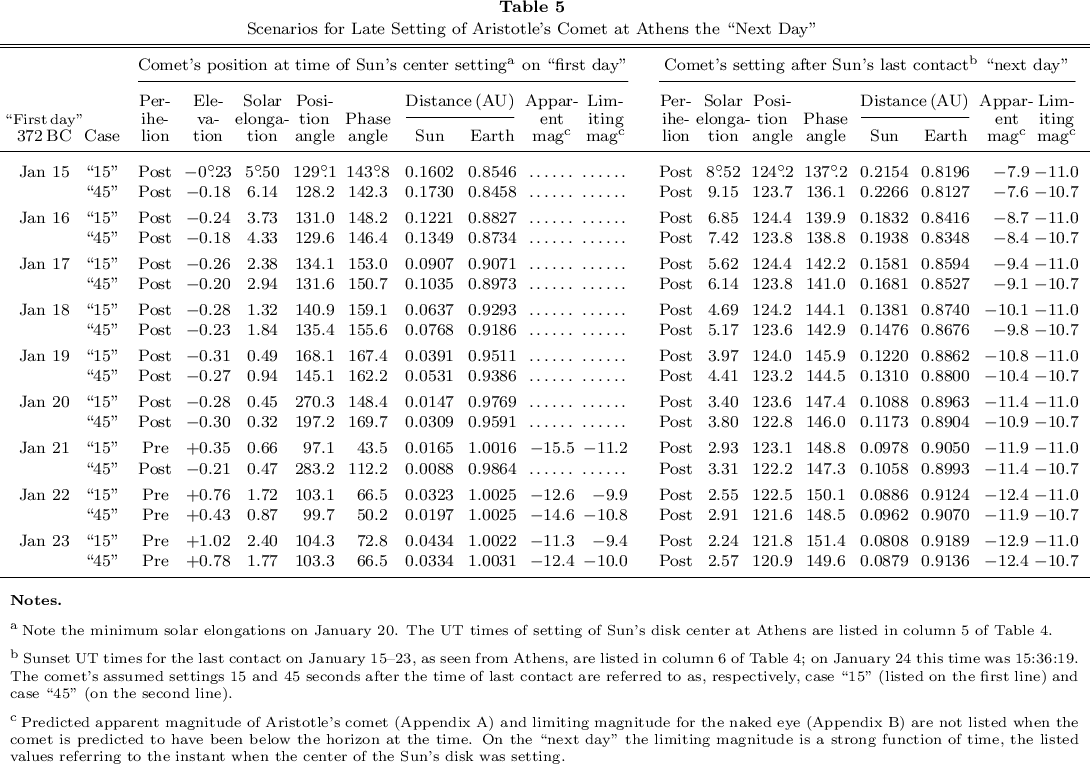}}}
\vspace{0.65cm}
\end{table*}

Additional information is presented in Tables~5 and 6.  Format of
Table~5 is similar to that of Table~3, except that the constraints
are now dictated by the comet setting after the Sun the ``next day''
rather than by the simultaneous settings of the comet and the Sun
on the ``first day''.  This information supplements the data in
Table~4 and explains its peculiar features.

The nice trend between the comet's settings and the sunset's first
contact between January~15 and 20, suddenly interrupted from
January~21 on, is shown in Table~5 to be the result of the case
``15'' and case ``45'' solutions swiching from post-perihelion
to preperihelion.  This table also shows the comet's solar
elongation at sunset to reach a minimum on January~20 and to
grow from 3$^\circ\!$.4 and 3$^\circ\!$.8 on January~21 (the
``next day'') to 8$^\circ\!$.5 and 9$^\circ\!$.2 on January~16
(the ``next day'' of January~15) in contradiction to Aristotle's
remark that the comet was very near the Sun.  In parallel with
increasing elongation, the comet's heliocentric distance was
increasing and the forward scattering effect decreasing.  All
these tests suggest that the scenario of January~20 in Tables~4
and 5 is the most attractive. 

The likelihood of January~20 as Aristotle's ``first day'' is
supported to an extent that it is fair to suggest that {\it the
young Aristotle's initial sighting of the comet was likely to have
happened on January~21, that is, the ``next~day'', which also was
the first day after the perihelion passage.}  For the sake of
argument, suppose he began to watch the comet 15~minutes before
it set on this day.  At the time it was 3$^\circ$--4$^\circ$ from
the Sun and less than 3$^\circ$ above the horizon, approaching
it at a rate of 0$^\circ\!$.175 per minute.  Showing the changes
in the limiting magnitude with time, Table~6 demonstrates that
the comet's head should easily have been perceived with the naked
eye almost all the way to the horizon, regardless of whether one
considers case ``15'' or case ``45'', in line with Aristotle's
narrative.

\begin{table}[b] 
\vspace{0.5cm}
\hspace{-0.2cm}
\centerline{
\scalebox{1}{
\includegraphics{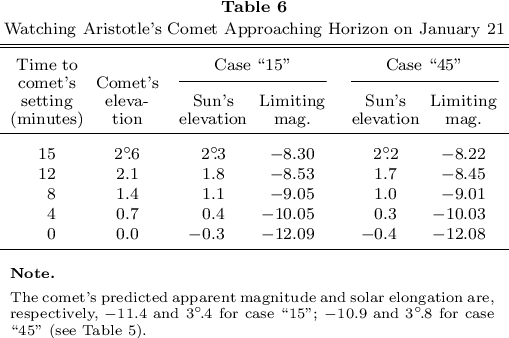}}}
\vspace{0.11cm}
\end{table}
\begin{table*}[t] 
\vspace{0.15cm}
\hspace{-0.18cm}
\centerline{
\scalebox{1.035}{
\includegraphics{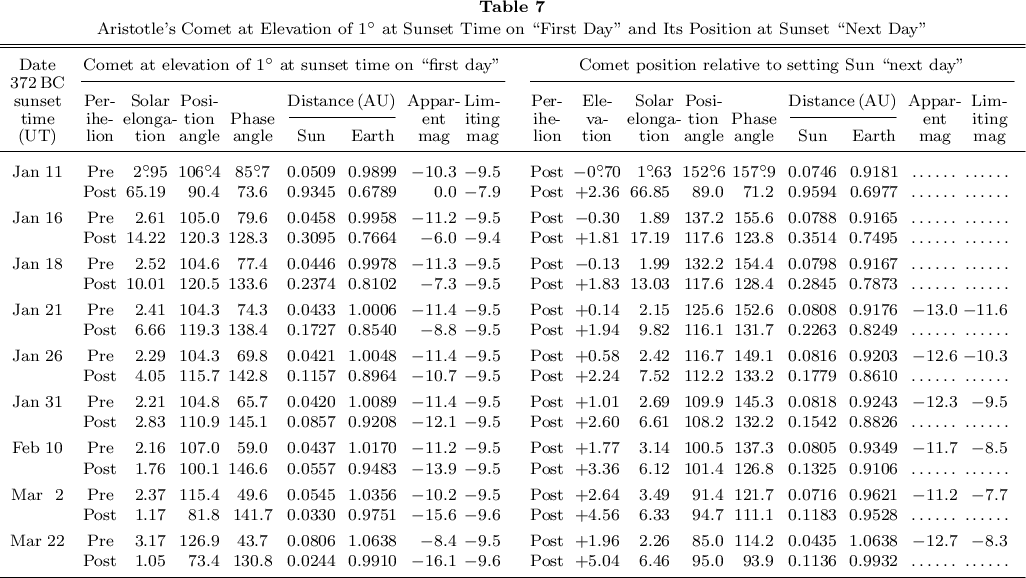}}}
\vspace{0.7cm}
\end{table*}

\subsection{Alternative Hypothesis Failing}
In the interest of offering complete analysis,~\mbox{Aristotle's} narrative
needs to be scrutinized from all angles.  A~hy\-pothesis that remains to
be examined is predicated on the prem\-ise that his remarks on
the comet setting before the Sun on the ``first day'' are incorrect
and that, instead, the comet set later than, yet too close to, the
Sun to be seen with the naked eye in its glare. 

Table~7 displays the results of such a solution in a case when
the comet's elevation at sunset on the ``first day'' was 1$^\circ$.
The results suggest that in the Preperihelion scenarios the comet
was, by and large, bright enough (except on March~22, an insupportable
outlier case) to be seen with the naked eye, and that the Post-perihelion
scenarios, which do show the comet to indeed have been too faint on the
``first day'' to be detected with the naked eye, place it much too
far from the Sun the ``next day''.

Similar additional random tests, performed for other assumed solar
elongations, failed to offer any positive evidence to question
Aristotle's remarks on the comet's whereabouts on the ``first day'',
suggesting that the conclusions presented in Section~5.3 stand.

\section{Motion of Aristotle's Comet and Orion's Belt}
The last sentence of Aristotle's narrative provides important
information on the comet's position just before it went out of
sight, which itself offers a crude estimate for the comet's
(or its tail's) brightness at the time.  It is fortunate that
the path happened to approach the famous asterism, as otherwise
Aristotle might not have at all mentioned the comet's location
in the sky.

In the following test of the predicted orbit of Aristotle's
comet, I compute a 5-day step topocentric ephemeris for Athens
(for equinox J2000) and plot the comet's post-perihelion path
across the sky as a function of perihelion time.  I consider
the same five dates, between January~1 and March~22, 372~BC,
which have been used in the preceding sections.  The results
are plotted in Figure~2 for the perihelion dates of January~1,
January~21, and February~10; in Figure~3 for the perihelion
dates of March~2 and March~22.  In each separate star map
position~1 refers to perihelion, position~2 to the time five
days later, position~3 another five days later, etc.  As the
positions in the plots get progressively more crowded with
increasing distance from perihelion, the step is increased
to 10~days after position~7, and sometimes to 20~days after
position~13.  For convenience, the relation between the
dates and the position numbers for each of the plots is
presented in Table~8.

\begin{figure*}[t] 
\vspace{-0.45cm}
\hspace{-0.2cm}
\centerline{
\scalebox{1.03}{
\includegraphics{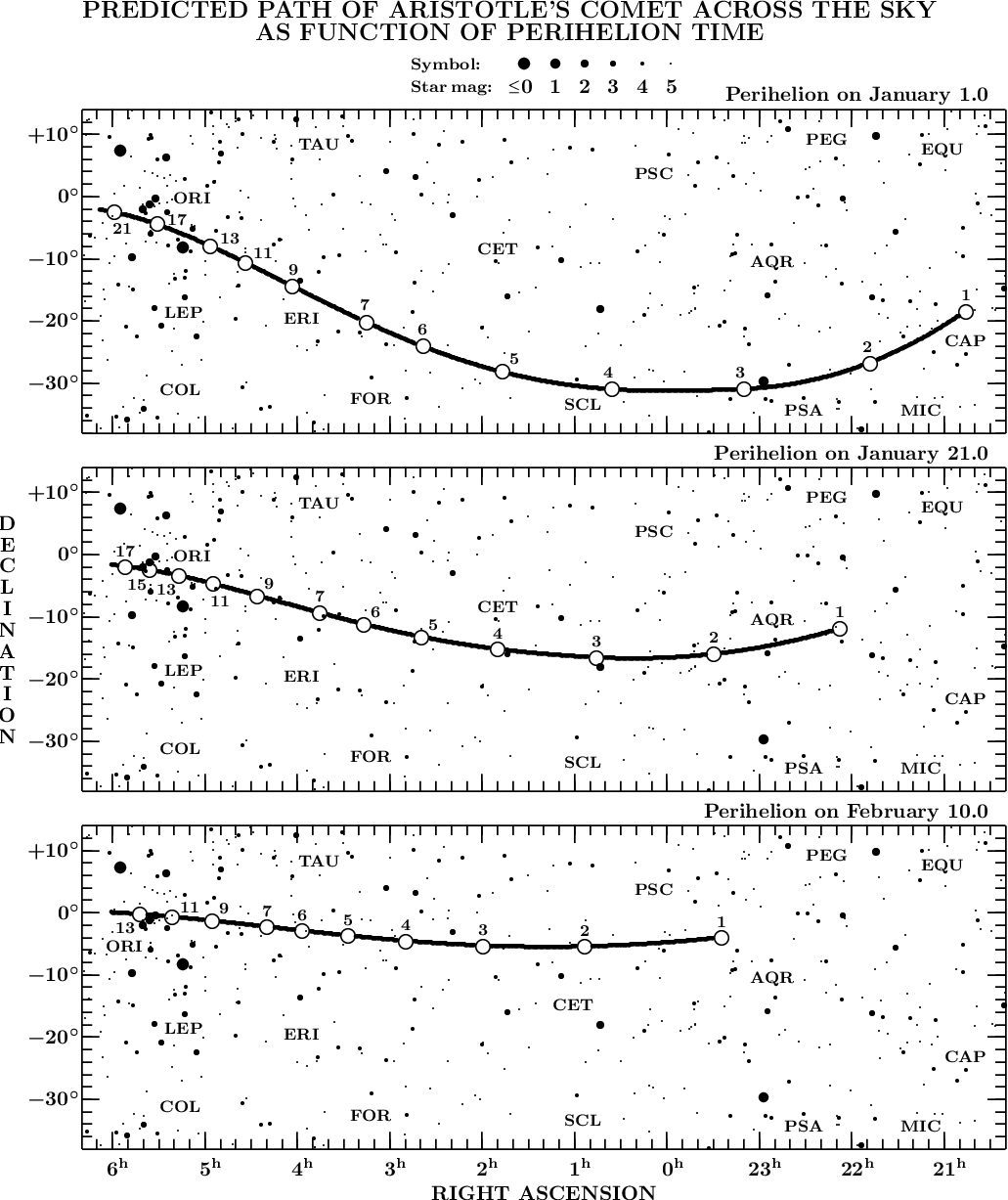}}}
\vspace{0cm}
\caption{Predicted post-perihelion path of Aristotle's comet as
seen from Athens in 372~BC (equinox J2000).  The star maps refer
to an assumed perihelion time at 0~UT on January~1 (top),
January~21 (middle), and February~10 (bottom).  The comet's
positions on standard dates are marked by large open circles
and position numbers that are explained in Table~8.  Position
number 1 always indicates the perihelion time.  The plots show that
if the comet passed perihelion on January~1, its path would have
missed Orion's belt to the south, but if it passed perihelion
between January~21 and February~10, the path would have crossed
the asterism between $\zeta$~Orionis (Alnitak) on April~3 in the
former case and $\delta$~Orionis (Mintaka) on April~4 in the
latter case.  Full moon was on January~1 and 30 and March~1 and
30.{\vspace{-0.5cm}}}  
\end{figure*}
\begin{figure*}[t] 
\vspace{0.2cm}
\hspace{-0.21cm}
\centerline{
\scalebox{1.08}{
\includegraphics{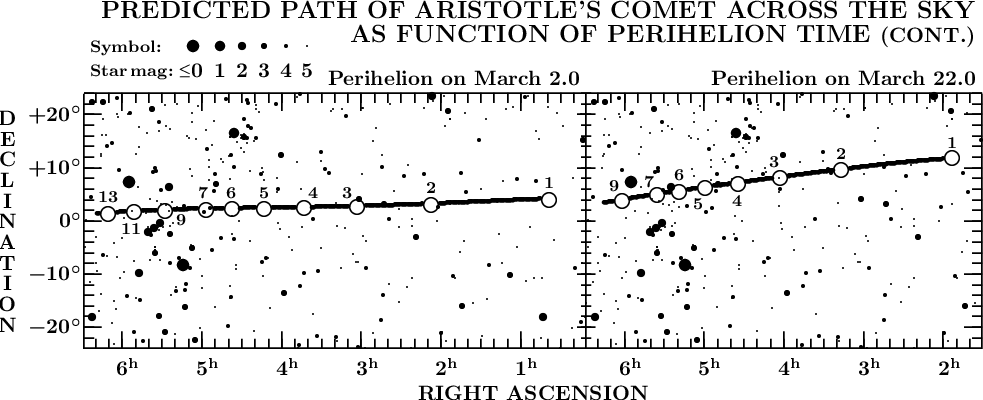}}}
\vspace{0cm}
\caption{Predicted post-perihelion path of Aristotle's comet as
seen from Athens in 372~BC (equinox J2000).  The star maps refer
to an assumed perihelion time at 0~UT on March~2 (left) and
March~22 (right).  As in Figure~2, the comet's positions on
standard dates are marked by large open circles and position
numbers (Table~8).  Again, position number 1 always indicates
the perihelion time.  The plot for the perihelion time of
March~2	shows the comet's path missing Orion's belt to the
north, while the path for the March~22 perihelion does no
longer even approach the belt stars.  Full moon was on March~1
and 30 and April~29.  Constellations are no longer
identified.{\nopagebreak}{\vspace{0.45cm}}}
\end{figure*}

In reference to Aristotle's remark that the ``comet receded as
far as Orion's belt,'' the plots in Figure~2 --- especially the
middle one --- are stunning.  They demonstrate that {\it
if the comet passed perihelion between about January~21 and
February~10, its predicted path did cross Orion's belt
on either April~3 or 4 --- near its southern boundary,
$\zeta$~Orionis, in the former case, near its northern boundary,
$\delta$~Orionis, in the latter case.  Here is another major
piece of evidence, which favors the applied set of orbital
elements, derived for Aristotle's comet as the Kreutz progenitor
in the framework of the contact-binary model and which is in
line with evidence examined in the previous sections of this
paper.\/}

By comparison, the comet's paths predicted for the perihelion times
of March~2 and March~22, presented in Figure~3, are disappointing.
The former misses the northern end of Orion's belt by a fair amount,
the latter does not get anywhere near the belt stars.{\vspace{-0.1cm}}

\begin{table}[b] 
\vspace{0.4cm}
\hspace{-0.16cm}
\centerline{
\scalebox{0.995}{
\includegraphics{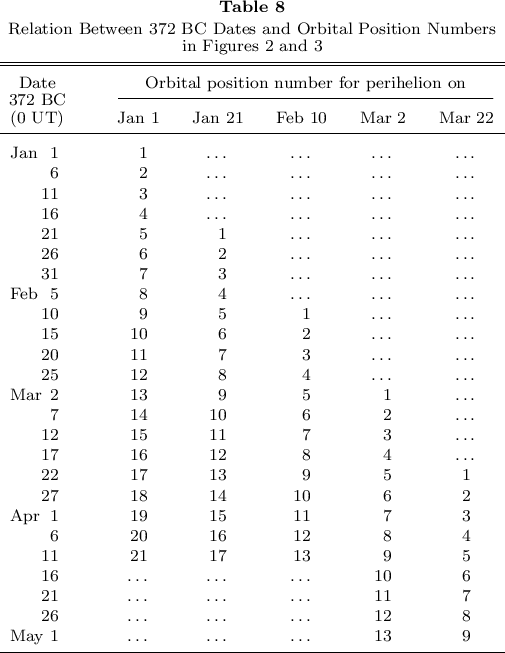}}}
\vspace{-0.05cm}
\end{table}

Of greatest interest is of course the scenario for the perihelion
time of January~21, which suggests that the path crossed Orion's
belt at close proximity of $\zeta$~Orionis late on April~3.
Combined with the time of first sighting on January~21
(Section~5.3), the comet appears to have been a naked-eye object
over a period of $\sim$72~days.  That is about one month longer than
both C/1843~D1 and X/1106~C1, but considerably less than C/1882~R1.
One might tentatively conclude that Aristotle's comet did not
fragment at perihelion as extensively as did the Great September
Comet of 1882.  And should it have arrived at perihelion around
February~10, rather than January~20, it would have been under
observation only a little over 50~days.

On the days around April 3 the conditions for sighting the comet
were not very good, even though there was no interference from
the Moon.  The comet was at a solar elongation of about 50$^\circ$.
Shortly after the beginning of astronomical twilight on April~3,
at 17:45~UT or 19:20 local time at Athens, the comet was at an
elevation of only about 18$^\circ$.  Thirty minutes later, at the
end of astronomical twilight, the elevation dropped to 12$^\circ\!$.5,
and the comet was setting another 65~minutes later.  As April was
progressing, the conditions further deteriorated.  It is conceivable
that the comet would have been visible a little longer, if the
conditions were more favorable.

\section{Tail Investigation}
The only piece of firm information offered by \mbox{Aristotle} on
the comet's tail is its length of 60$^\circ$, as it ``extended
over a third part of the sky.''  In addition, one version of the
translated text could possibly be understood to indicate that the
tail stretched to Orion's belt, although this is unlikely.  Since
Aristotle provides no time reference, it might nonetheless be
interesting to estimate the position in the sky that the comet
would have to occupy in order that the end of its 60$^\circ$
long tail reach Orion's belt.

\begin{figure}[t] 
\vspace{0.86cm}
\hspace{-0.535cm}
\centerline{
\scalebox{1.375}{
\includegraphics{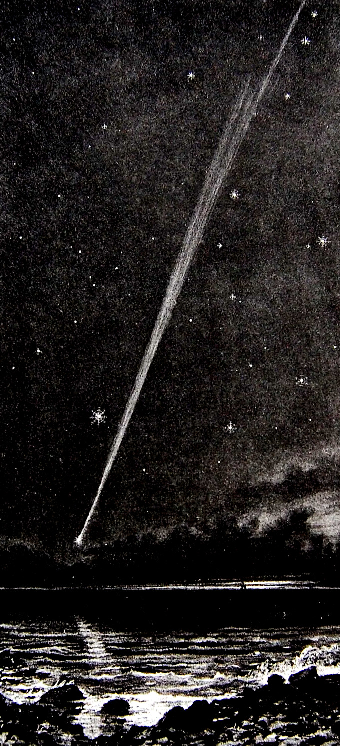}}}
\vspace{0.05cm}
\caption{The Great March Comet of 1843 in the constellation of Cetus
seen from Cape Town in the evening of March~4, 1843 shortly before it
was setting at 18:36~UT.  The comet was 4.86~days after perihelion at
a solar elongation of about 17$^\circ$.  According to his diary,
C. Piazzi Smyth observed the comet near the Signal~Station on the
Lion's Rump, from where it was seen to set on the~sea~horizon (Warner
1980).  The painting was apparently made from a location on a nearby
beach.  Although the diary provides no information on the tail length,
from the configuration of the bright stars $\beta$, $\eta$, $\theta$,
$\zeta$, and $\tau$~Ceti, depicted in the picture, one can determine
a length of $\sim$34$^\circ$, which includes the streamer at the upper
end.  The tail's axis pointed almost exactly in the antisolar direction,
at a position angle of 106$^\circ$.  About 0.5~AU long in space, the
tail was essentially pure plasma, but microscopic dust grains may have
contributed a little near the head.  Farther from the head the
tail became double, the distance between the two rays slightly
increasing toward the upper end.  The painting shows the comet's
reflection in water. (Detail of painting; National Maritime
Museum,~London.){\vspace{-0.3cm}}}
\end{figure}

The condition to be satisfied in this scenario is
\begin{equation}
\cos 60^\circ\! = \sin \delta \sin \delta_{\rm belt} + \cos \delta
  \cos \delta_{\rm belt} \cos(\alpha_{\rm belt} \!-\! \alpha),
\end{equation}
where $\alpha$ and $\delta$ are, respectively, the right ascension
and declination of the comet, and $\alpha_{\rm belt}$ and
$\delta_{\rm belt}$ are the representative equatorial coordinates
of Orion's belt.  Equating them, for example, with those
of the belt's middle star, $\epsilon$ Orionis (Alnilam), one gets
for the comet's position a relation (equinox J2000):
\begin{equation}
\alpha = 84^\circ\!.05 - \arccos (0.50 \sec \delta + 0.02 \tan \delta).
\end{equation}
This condition requires that \mbox{$\alpha = 1^{\rm h}42^{\rm m\!\!}.82$}
for \mbox{$\delta = -20^\circ$} and \mbox{$\alpha = 1^{\rm h}39^{\rm
m\!\!}.46$} for \mbox{$\delta = -15^\circ$}.  Interpolation and
comparison with a topocentric ephemeris of the comet suggests that
the condition of a 60$^\circ$ long tail terminating at Orion's belt
would have been satisfied on February~4, or 15 days after the
perihelion on January~20.  However, there is a problem with the
tail's orientation.  The position angle of Orion's belt from the
comet~was~at~the~time 82$^\circ \pm$\,1$^\circ$, while the position
angle of the tail extending in the antisolar direction,
was 99$^\circ$, nearly 20$^\circ$ off.  Because the inferred
time was merely 1.5~days before the Earth's predicted crossing of
the comet's orbit plane, the orientations of the plasma and dust
tail essentially coincided.  Accordingly, the discrepancy is well
established and cannot be removed, given the adopted set of
orbital elements in Table~1.  However, when one
accepts Webster's (2004) translation of Aristotle's text on
Orion's belt as a reference to the position of the comet just
before its disappearance, rather than its tail's termination point,
this controversy of course becomes irrelevant.

\subsection{``Path'' Tail (Plasma)}
An important issue is {\it when\/} did the tail reach the length of
60$^\circ$?  Aristotle does not answer this question unequivocally.
Can one rule out the possibility that it happened the ``next day'',
the first evening after perihelion?  Even if the tail attained this
length a few days later --- a more plausible scenario --- it must
have been of the plasma variety, as dust could not possibly be
accelerated to such enormous distances from the nucleus in such a
short period of time.  An excellent example is the Great March Comet
of 1843 (C/1843~D1), a third-generation fragment of Aristotle's comet,
which less than five days after perihelion is known to have displayed
a plasma tail 34$^\circ$ in length, as illustrated in Figure~4.

\begin{table}[b] 
\vspace{0.45cm}
\hspace{-0.2cm}
\centerline{
\scalebox{1}{
\includegraphics{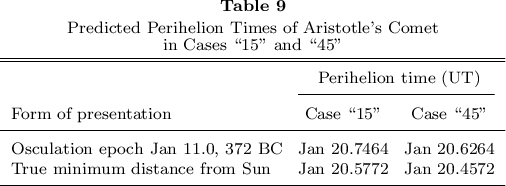}}}
\vspace{0.04cm}
\end{table}

While it is not surprising that the tail of Aristotle's comet was
nearly twice as long, to make a reconstruction of the comet's appearance
compelling requires a comprehensive investigation.  I revert to
the case~``15'' and case~``45'' scenarios introduced in Section~5.3
and present the predicted perihelion times in Table~9.  They are used
to list, in Table~10, the comet's elevation, solar elongation, phase
angle, position angle of the radius vector, and other relevant data
for either scenario at sunset on a number of days starting with
January~21, identified in Section~5.3 as Aristotle's ``next day''.
The linear length of the 60$^\circ$ plasma tail extending along the
radius vector is plotted in Figure~5 as a function of observation
time.\,\,\,

\begin{table*}[t] 
\vspace{0.15cm}
\hspace{-0.2cm}
\centerline{
\scalebox{0.98}{
\includegraphics{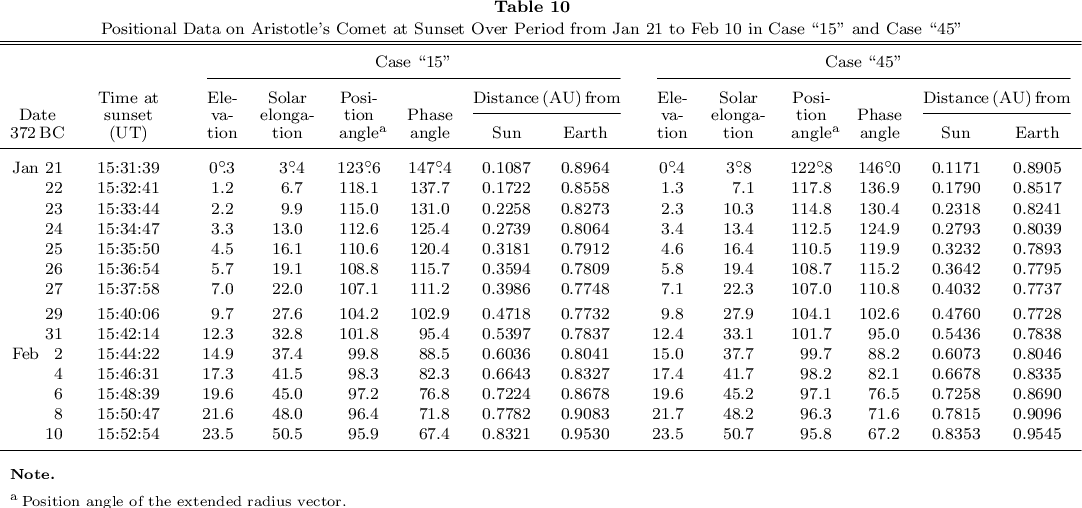}}}
\vspace{0.5cm}
\end{table*}

Table 10 shows that there is very little difference in the parameters
between case~``15'' and case~``45''.  In particular, the position angle
in which the plasma tail is predicted to have been pointing at sunset
on January~21 is in either case near 123$^\circ$, in a direction above
the horizon, as will be shown below.  At sunset on the following days
the position angle was decreasing at a fairly high rate, thereby
improving the tail's observing conditions.  Also helpful were the comet's
gradually increasing elevation and solar elongation.

\begin{figure}[b] 
\vspace{0.5cm}
\hspace{-0.2cm}
\centerline{
\scalebox{0.77}{
\includegraphics{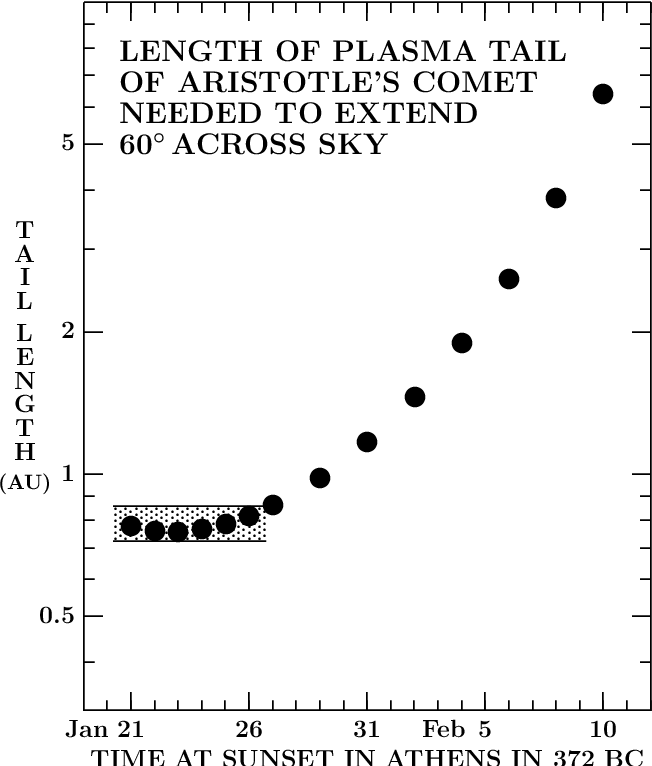}}}
\vspace{0cm}
\caption{Length of the plasma tail of Aristotle's comet in
space, predicted on the assumption that the tail extends along
the radius vector, vs an (unknown) time at which its
observed length was 60$^\circ$ (or ``a third part of the sky'')
according to Aristotle.  The dates of January~21 through January~26
(shaded area) consistently suggest a tail length of $\sim$0.8~AU.
From the beginning of February on the lengths become much too large,
increasing~exponentially.{\vspace{-0.08cm}}}
\end{figure}

On the assumption that the tail extends along the radius vector away
from the Sun, its observed length, $\ell$, is converted to its true
length in space, $\Lambda$, according to a well-known formula
\begin{equation}
\Lambda = \frac{\Delta \sin \ell}{\sin(\psi \!-\! \ell)},
\end{equation}
where $\Delta$ is the distance of the comet's nucleus from the observer
and $\psi$ is the phase angle.  Using the data from Table~10, the plasma
tail's predicted length, corresponding to the observed length of
60$^\circ$, is shown in Figure~5 to have equaled about 0.8~AU, if
it was seen between January~21 and 26.  Starting with January~27,
the length needed to fit 60$^\circ$ began to grow exponentially,
reaching a meaningless range from 1.9~AU on February~4 to 6.4~AU
on February~10.  It is obvious from Equation~(4) that to avoid
unacceptable solutions, the observed length cannot approach the
phase angle.  Yet, Table~10 shows that on February~10 the phase angle
diminished to 67$^\circ$, suggesting that on this date the tail must
have been considerably shorter than 60$^\circ$, possibly shorter than
35$^\circ$.  Alternatively, it deviated significantly from the radius
vector.  However, for plasma tails this latter possibility is not
corroborated by evidence based on observations of other spectacular
Kreutz sungrazers.

I consider it likely that Aristotle's reference to the tail's length
applies to a date before the end of January and that the tail was
not much longer than 0.8~AU.  Comparison with the three brilliant
sungrazers of the 19th and 20th centuries (C/1843~D1, C/1882~R1,
and C/1965~S1) supports this conclusion, as their typical tail
lengths were \mbox{0.32--0.55}~AU 5~days after perihelion and
\mbox{0.60--0.83}~AU 15~days after perihelion (Sekanina 2023).

\begin{table}[t] 
\vspace{0.15cm}
\hspace{-0.17cm}
\centerline{
\scalebox{0.98}{
\includegraphics{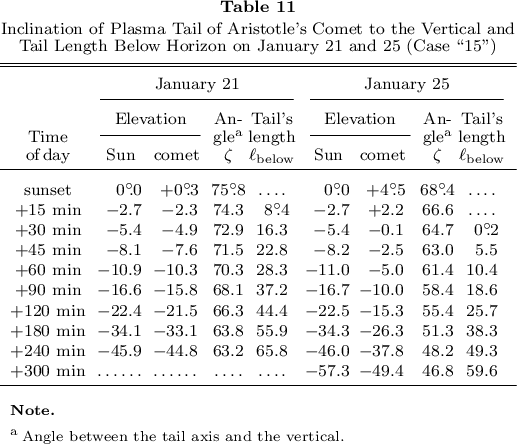}}}
\vspace{0.45cm}
\end{table}

Next I address the issues of an apparent orientation of the observed
tail relative to the horizon and, after the comet's setting, the
length of the tail hidden below the horizon.  Even if the tail
was rather prominent, only its brighter parts could at best be
seen at sunset.  If the comet was ``ever so little behind the
Sun and setting immediately'' on January~21, it was not possible
to estimate the total length of the tail in the dark sky, because
much of it was setting soon after the Sun.  How much, depends
on the angle $\zeta$ that the tail made at the comet's nucleus
with the vertical to the horizon.  Angle $\zeta$ is conveniently
computed as a difference between the position angle of the tail,
$P_{\rm tail}$, assumed to coincide with the position angle of
the extended radius vector, and the position angle of the zenith,
$P_{\rm znth}$, given by Equations~(1) but with the Sun's angles
now replaced by those of the comet's nucleus,
\begin{equation}
\zeta = P_{\rm tail} \!-\! P_{\rm znth}.
\end{equation}
At the comet head's depression of \mbox{$\eta = -h$} (where $h$ is
the elevation, \mbox{$h \!>\! 0$} when the comet is above the horizon
and vice versa), the fraction of the tail's length below the horizon,
$\ell_{\rm below}$, equals
\begin{equation}
\ell_{\rm below} = \arctan \!\left(\!\frac{\tan \eta}{\cos \zeta}\!\right).
\end{equation}

Since the numbers in Table 10 for case ``45'' differ very little from
those for case ``15'', all computations below will below be performed
only with the parameters of the latter case.  The zenithal position angle
is found to have varied from 48$^\circ$ on January~21 to 42$^\circ$ on
January~25, to 35$^\circ$ on January~29, and to 29$^\circ$ on February~2.
Accordingly, the tail is predicted to have made an angle of 76$^\circ$
with the vertical to the horizon on January~21, dropping to a minimum
of 68$^\circ$ on January~25 and the two following days, and then
increasing slowly to 69$^\circ$ and 71$^\circ$ on, respectively,
January~29 and February~2.  By that time the comet was at sunset
already 37$^\circ$ from the Sun and at an elevation of 15$^\circ$
(Table~10).  The tail appears to have been deviating significantly
enough from the vertical at sunset that its length above the horizon
shortened very rapidly after sunset, especially on January~21.

I now examine how did the inclination of the tail to the horizon
--- or the complementary angle $\zeta$ --- affect its length that
the observer saw after the comet had set.  This angle and the part
of the tail that was hidden below the horizon after the comet's
setting are presented in Table~11 for January~21 and 25 as a
function of the time after sunset.  The difference between the
two evenings, only four days apart, was enormous.  At the end of
civil twilight, a little over half an hour after sunset, nearly
20$^\circ$ of the tail was already below the horizon, so that
a tail that would be seen to extend over 60$^\circ$ was in
its entirety almost 80$^\circ$ long.  On the other hand, at the
same time on January~25 practically the whole tail was still
above the horizon!  A similar disparity is shown in the table
at the other times; for example, at the beginning of astronomical
twilight, longer than one hour after sunset, one half of a
60$^\circ$ long tail would have already beem below the horizon
on January~21, but only about one fifth of it on January~25.

The last, very crude plasma-tail test that I undertake is the
computation of an average radial acceleration that the charged
particles would need to experience in order to reach a distance
of 0.8~AU from the nucleus in the given period of time.
I find that in case ``15'' the ions released at or very shortly
after perihelion would have had to be under an average acceleration
of at least 6~times the solar gravitational acceleration to reach
a distance of 0.8~AU from the nucleus just after sunset on
January~21, or at least 10~times the solar gravitational acceleration
to reach a distance of 1~AU.  For the tail observed on January~25,
the ion acceleration was not an issue.

The presented arguments suggest that Aristotle's~com\-et did not in all
probability display its majestic, 60$^\circ$ long tail on January~21
(the ``next day''), but days later.{\vspace{-0.13cm}}

\subsection{Dust Tail}
The general conditions for the dust-tail formation and evolution
in spectacular sungrazing comets were addressed in a recent paper
(Sekanina 2025) and will not be repeated here.  Constraining the
radiation-pressure accelerations to \mbox{$\beta \leq 0.6$} the
solar gravitational acceleration, to which particles were subjected
in the dust tails of other Kreutz sungrazers, the dust tail of
Aristotle's comet on January~23.65 and 25.65~UT, shortly after
sunset, is schematically modeled by syndynames in Figure~6,
reflecting a setup in case ``15''.

The profound curvature of the tail's predicted contours, plainly
seen especially on January~23.65~UT, is utterly incompatible with
the ``path''-like appearance of the observed tail, remarked on
by Aristotle.  However, also notable in Figure~6 is the dramatic
lengthening as well as straightening of the dust tail over the
pariod of mere 48~hours, between January~23 and 25.  It appears
that dust contributed very little to the observed tail in the early
post-perihelion days.  This situation was gradually changing with
the comet's increasing heliocentric distance on the following
days.  Affecting the appearance of the dust tail was also the
Earth's transit across the orbit's orbital plane, which occurred
on February~5.

\begin{figure}[t] 
\vspace{0.17cm}
\hspace{-0.2cm}
\centerline{
\scalebox{0.73}{
\includegraphics{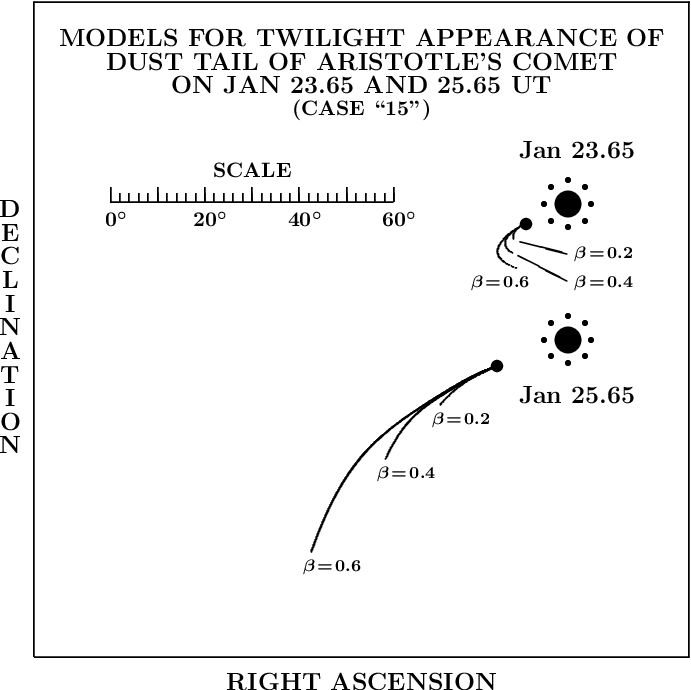}}}
\vspace{0cm}
\caption{The modeled schematic appearance of the dust tail of
Aristotle's comet on January~23.65 and 25.65 UT, shortly after
sunset, in case ``15'' (equinox J2000).  The tail contours are
depicted by three syndynames, showing particle trajectories
affected by the radiation-pressure accelerations of, respectively,
\mbox{$\beta = 0.2$}, 0.4, and 0.6 the solar gravitational
acceleration.  The syndynames show ejecta starting one hour
after perihelion.  The dramatic expansion and straightening
of the tail contours is plainly seen.  Yet, in this period
of time the contribution from dust to the observed tail,
essentially a plasma feature and described by Aristotle as
a ``path'', was limited to a region near the comet's
head.{\vspace{0.5cm}}}
\end{figure}

Systematic changes in the properties of the dust tail over a
period of 20~days after perihelion are described in Table~12 in
positional terms of a particle subjected to a radiation-pressure
acceleration of 0.6 the solar gravitational acceleration (a
submicron-sized silicate grain).  The table shows, as a function
of the particle's ejection time, its distances from the Sun and
Earth, apparent distance from the comet's nucleus (as the tail's
length), position angle, and phase angle at the time of observation.

\begin{table}[t] 
\vspace{0.15cm}
\hspace{-0.2cm}
\centerline{
\scalebox{0.988}{
\includegraphics{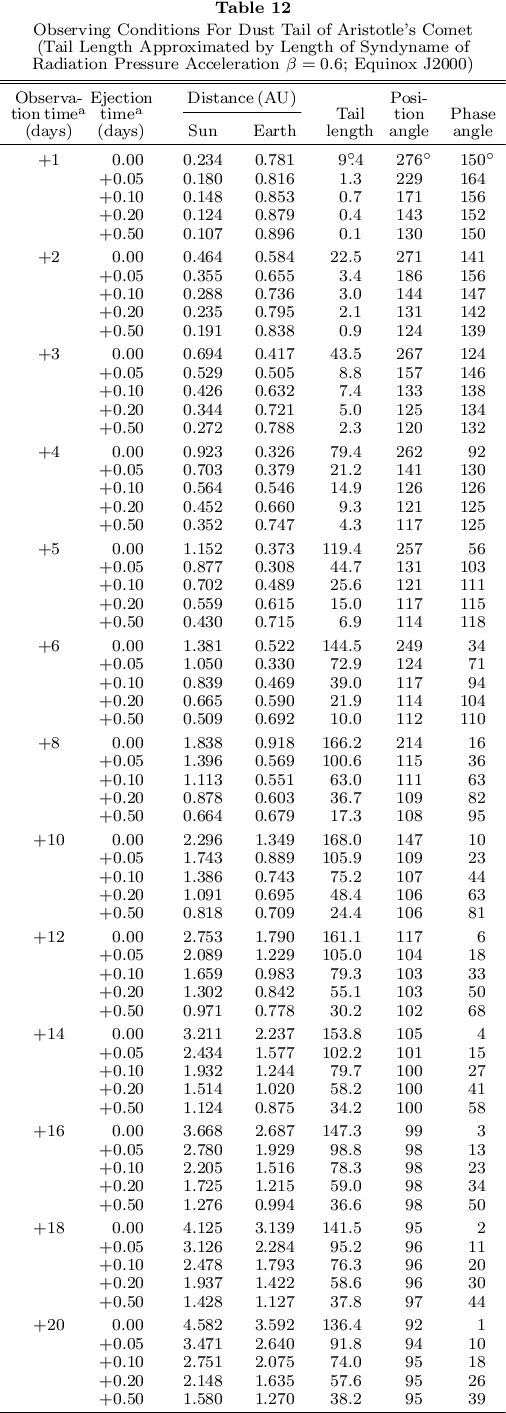}}}
\vspace{-0.2cm}
\end{table}

As the observed tail was oriented essentially along the extended
radius vector, at position angles in a range of 100$^\circ$ to
120$^\circ$, the dust ejecta released at perihelion and very
shortly after perihelion appear to have never contributed to
the observed tail.  At first sight this looks strange, but the
experience with the SOHO sungrazers shows that their last
astrometric measurement in the C2 coronagraph is possible,
on the average, 0.26~day before perihelion, for the brightest
objects at about 0.12~day before perihelion.  At those times
these heavily fragmented objects disintegrate by sublimation.
The implication is that much of the dust ejected at
post-perihelion times less than, say, 0.1~day after perihelion
is likewise subject to sublimation, so these grains do not
survive.

Interestingly, if these presumably nonsurviving dust particles
are eliminated, Table~12 suggests that the dust tail could have
reached a length of $\sim$60$^\circ$ in the expected antisolar
direction approximately 8~days after perihelion (or January~28),
if one takes a cutoff ejection time at 0.1~day; or 16~days after
perihelion (or February~5) for a cutoff at 0.2~day.\\[0.2cm]
\_\_\_\_\_\_\_\_\_\_\_\_\_\_\_\_\_\_\_ \\
\noindent
{\scriptsize \bf Note to Table 12}\\
\noindent
{\scriptsize $^{\rm a}$\,Reckoned from perihelion; plus sign
means a post-perihelion time.}

The situation was different around the time of the last sightings
of the comet, which according to Aristotle was then located at
Orion's belt.  At a heliocentric distance of 1.9~AU, a geocentric
distance of 2.4~AU, and a solar elongation of 49$^\circ$, the
plasma tail was long gone and the outer regions of the dust tail
scattered in space to the extent that they were well below the
detection threshold of the naked eye.  It could reasonably be
expected that the last sightings of the comet were those of the
brightest dust-tail areas made up of larger particles subjected
to radiation-pressure accelerations of much less than 0.1 the
solar gravitational acceleration, which would imply a tail several
degrees long on April~3, 372~BC, seen in Athens at the end of
astronomical twilight (at about 18:15~UT).  The predicted position
of the comet and its tail at that time are schematically shown in
Figure~7.  If the dates inferred in this paper from Aristotle's
description are correct, the comet was seen for longer than 10 weeks.

\section{Ephemeris}
For reference, I present in Table 13 the comet's geocentric ephemeris
at a step of five days. It has been computed from the orbital elements
listed in the right column of Table~1 and the perihelion time for
case ``15'' shown in Table~9.  The magnitudes have been derived from
the formulas for the adopted light curve in Appendix A.

\begin{figure}[b] 
\vspace{0.65cm}
\hspace{-0.2cm}
\centerline{
\scalebox{0.765}{
\includegraphics{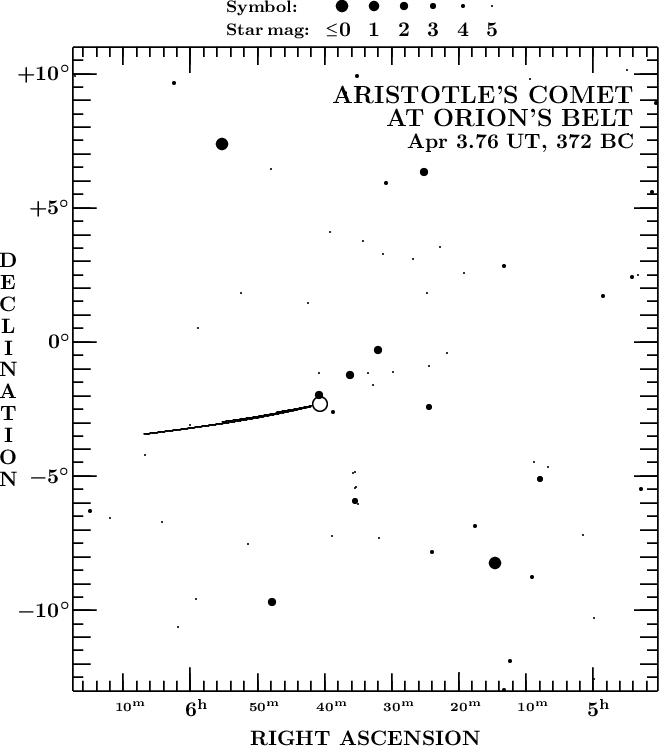}}}
\vspace{0cm}
\caption{Aristotle's comet about the time of its final sighting at Orion's
belt, as it may have appeared on April 3.76 UT, 372~BC, at the end of
astronomical twilight at Athens.  Its elevation was then 12$^\circ\!$.5.
Its head (large open circle), whose predicted location was less than
22$^\prime$ to the south of $\zeta$~Orionis, at the southeastern end
of Orion's belt, is likely to have possessed a residual dust tail
several degrees in length, pointing at a position angle of approximately
100$^\circ$ and composed of particles subjected to radiation-pressure
accelerations lower than 0.1 the solar gravitational acceleration.
The head was almost certainly not visible with the naked eye~any~longer.
The map's equinox is J2000.{\vspace{-0.1cm}}} 
\end{figure}
\begin{table}[t] 
\vspace{0.15cm}
\hspace{-0.16cm}
\centerline{
\scalebox{0.997}{
\includegraphics{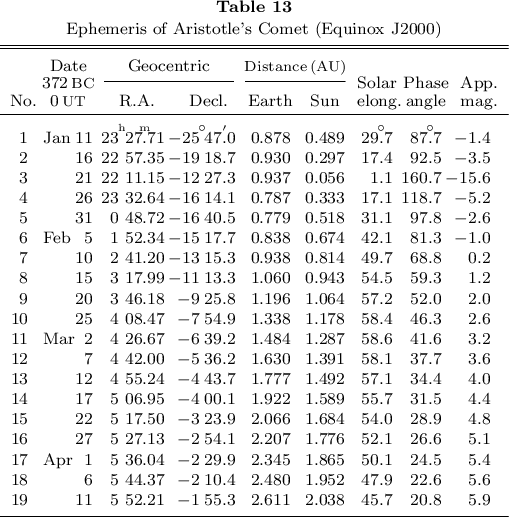}}}
\vspace{0.5cm}
\end{table}

\section{Final Comments and Conclusions}
The prime objective of this paper has been to exploit the opportunity
provided by Sekanina \& Kracht's (2022) determination of the orbital
elements of Aristotle's comet in the process of testing the new
contact-binary model for the Kreutz sungrazer system (Sekanina 2021).
The only element that the verification process was unable to determine
with satisfactory accuracy was the perihelion time, which could be
estimated to, say, $\pm$1~month or perhaps even slightly better (Sekanina
2022a).  There was a chance that by confronting the predicted orbit
with Aristotle's remarks on the comet in his treatise {\it
Meteorologica\/} could lead to an improved estimate of the perihelion
time.

Aristotle's treatise must have been, over the ages, translated into
English many times, and one would expect that the translated text was
always more or less identical.  Comparison of two such products does
however show that this is not the case.  In this study I have relied
heavily on the translation by Webster (2004).

One can only speculate what thoughts and experience led Aristotle to
put down the information on the comet with the words he used.  However,
the text leaves the impression that he saw the comet, as a youngster,
himself.  Although he does not mention any dates, he does identify
the year (twice:\ by the name of the Athens archon and by remarking
on the Helike earthquake, which is dated accurately) and the season
(winter and frosty weather).  The statement that one day ``the comet
was not seen, as it set before the Sun'' may be perceived as somewhat
speculative, but a numerical test shows its surprisingly constraning
impact.  I assume that the ``next day'' means 24~hours later, but
when did the tail extend ``over a third part of the sky'' is unclear.
My educated guess is that for days, starting less than a week
following~the~``next~day''.  Aristotle's last sentence is unusual in
that in the two versions that I compared its translation is different.
In Webster's version the meaning is unequivocal, offering important
positional and brightness constraints.

Given that the account of the comet was written by Aristotle some
40~or so years after experiencing the event, his description
contains some remarkable details, suggesting that his comments on
the comet were a synopsis of his thoughts at the time of writing
and notes on the comet's appearance, position, and motion that
he was collecting over time for this very purpose.  Aristotle
was a notable philosopher and the style of his writing differs
dramatically from that of some of the ancient historians, such as
Diodorus Siculus, whose description of the comet written about
three centuries later is utterly useless.

Most calculations in this paper were made on the assumption that
the comet's observer was in Athens.  It is known that Aristotle
was born in the city of Stagira in the peninsula of Chalkidiki,
about 290~km north of Athens.  Where he lived in 372~BC is unclear;
his parents died early and his guardian became Proxenus, who was
a resident of Atarneus, a city about 110~km southeast of Troy.
Some time in his childhood Aristotle may have spent in Pella,
the Macedonian capital, not far from his birthplace.  At the age
of 17 or 18, he moved to Plato's Academy in Athens.  Since it
is likely that the comet was one of many topics that Aristotle
debated with other educated Greeks in the Academy and elsewhere,
his account in {\it Meteorologica\/} is perhaps based as much on
his personal observations as on collective knowledge that he had
been gradually gaining.  This knowledge was acquired at different,
presumably mostly Greek locations, so that the ``observing site''
is indeterminate except that it was in Greece.  And since the
observing site plays a role in this paper mainly when a relative
position of the comet to the Sun is at issue, Athens is as
representative a location for the observer as any.

I have used Aristotle's narrative to examine three conditions on
the comet's perihelion time.  The requirement of the comet setting
before the Sun one day and just after the Sun the following day is
highly restrictive.  It leads to January~20 as the most probable
perihelion date, with two closely investigated cases on January~21,
in which the comet was assumed to set, respectively, 15 or 45~seconds
later than the Sun (cases ``15'' and ``45''), resulting in the perihelion
times of less than 3~hours apart.  This solution essentially means that
on Aristotle's ``first day'' the comet was not seen because it was either
behind the Sun's disk or, perhaps more probably, in near contact with it
in the sky.\footnote{The reader may wonder why the elaborate examination,
when this conclusion is apparent from cursory inspection of Figure~1.
The reasons why Figure 1 does not suffice are:\ (i)~the path of the
comet relative to the Sun is plotted in the equatorial, rather than
horizontal, coordinate system; (ii)~the relative positions of the
comet at perihelion and the Sun are very similar on January~1 and 21,
yet the January~1 solution is found in the paper to be unacceptable;
and (iii)~the figure offers no information on the solutions, closely
examined in the paper, that explore chances that perihelion occurred
on a day other than one of the five standard dates.}  Aristotle's
condition for the comet's setting the ``next day'' immediately after
sunset makes it certain that the comet was then very probably just
about one day after perihelion.

The comet's orbit with the perihelion taking place~on January~20
(case ``15'') is found to have passed 21$^\prime\!$.6~from
$\zeta$~Orionis, the southernmost bright star of Orion's belt,
on April~3.92~UT.  The comet's crossing the belt near
$\delta$~Orionis, the northernmost bright star, requires that the
perihelion have occurred on February~10.  If by ``receding as far
as Orion's belt'' did Aristotle mean crossing the belt essentially
between the two bright end-belt stars, the range of perihelion times
allowed by this constraint is a three-week interval between January~20
and February~10, a less tight condition than the previous one.

The third condition is a 60$^\circ$ long tail; it must have been
dominated by plasma, just as in the painting of the Great March
Comet of 1843 made about 5 days after perihelion.  If the tail
that Aristotle refers to lined up along the radius
vector, its linear length in space would have been 0.8~AU if
seen between January~21 (highly unlikely) and January~26.  Its
length would have grown exponentially if seen at later times,
reaching an improbable length of 1.9~AU if seen on February~5 and
an absurd length of 6.4~AU if seen on February~10.  I estimate that
the most likely time of the plasma tail reaching its projected
length of 60$^\circ$ was several days after perihelion, around
January~25, when the comet was at a heliocentric distance of
about 0.3~AU.

The contribution from dust to the tail was minimal shortly after
perihelion, but was growing gradually farther from the Sun, as
the plasma tail began to weaken.  Dust eventually took over
completely, and just before the comet disappeared, it was a
several-degree long dust tail in position angle of about
100$^\circ$ that the observers saw at Orion's belt.

This summary completes the present study, in which a set of
orbital elements of the Great Comet, predicated on the recently
proposed contact-binary model, was confronted with Aristotle's eyewitness
remarks.  The results strengthen the notion that this magnificent
object indeed was the progenitor of Kreutz sungrazers, passing
perihelion most probably on January 20, 372~BC and visible to the
naked~eye over a period of more than 10~weeks. \\[-0.2cm]

%
\begin{center}
{\large \bf Appendix A} \\[0.3cm]
THE ADOPTED LIGHT CURVE OF ARISTOTLE'S~COMET \\[0.3cm]
\end{center}
In order to examine the visibility of Aristotle's comet in the
various positions near the Sun, one needs to have some idea
about its observed brightness, which near the Sun varies
rapidly with time as well as with the comet's position in
space.

In the absence of any information I use a limited set of data
on other Kreutz sungrazers to {\it adopt\/} a light curve for
Aristotle's comet.  Its observed brightness is investigated
in terms of an apparent magnitude, $H_{\rm app}$, which is
assumed to depend on the distances from the Sun, $r$, and
Earth, $\Delta$, as well as on a function $\Phi(\psi)$ of the
phase angle, $\psi$.  For $\Phi(\psi)$ I adopt the law developed
by Marcus (2007) for dusty comets.  In addition, the light curve
depends on two constant parameters that are intrinsic to the
comet:\ (i)~the absolute magnitude, $H_0$, which is the magnitude
that the comet would have at unit heliocentric and geocentric
distances \mbox{$r = \Delta = 1$ AU} and a zero phase angle,
\mbox{$\psi = 0$}, for which the phase function is zero,
\mbox{$\Phi(0^\circ) = 0$}; and (ii)~the inverse power of
heliocentric distance, $n$, with which the comet's brightness
varies, $r^{-n}$.  The expression for the apparent magnitude is
\begin{eqnarray*}
\,H_{\rm app}(r,\Delta,\psi;H_0,n) & = & H_0 + 2.5 n \log r +
 5 \log \Delta + \Phi(\psi), \\[-0.05cm]
	& & {\hspace{4.28cm}}\mbox{\rm (A-1)} \\[-0.8cm]
\end{eqnarray*}
where \mbox{$H_{\rm app}(1,1,0^\circ;H_0,n) = H_0$} for arbitrary $n$.

Since the light curves of bright Kreutz sungrazers are known to be,
in general, asymmetric relative to perihelion, there are separate
preperihelion{\vspace{-0.055cm}} parameters, $H_0^-$, $n^-$, and
post-perihelion parameters, $H_0^+$, $n^+$.  The most complete light
curve, both before and after perihelion, is available for comet
Ikeya-Seki (C/1965~S1),{\vspace{-0.04cm}} which is fitted with
\mbox{$H_0^- = H_0^+ = 5.9$} and \mbox{$n^- = n^+ = 4.0$} (Sekanina
2002).  As data on the preperihelion light curves of other bright
sungrazers are very fragmentary, in the following I will
assume that Aristotle's comet brightened at the same rate,
\begin{displaymath}
n^- = 4.0. \rlap{\hspace{2.77cm}\mbox{\rm (A-2)}}
\end{displaymath}

After perihelion, Ikeya-Seki is known to display two persistent
nuclei.  On the other hand, the Great September Comet of
1882 was fading less steeply, varying as $r^{-3.3}$.  As is
well known, this sungrazer fragmented heavily at perihelion,
exhibiting five to six nuclei when receding from the Sun.
Finally, there is some evidence that the Great March Comet
of 1843, which was not observed to split, was fading with
a power of heliocentric distance \mbox{$n^+ > 4$}.  Based
on this very limited amount of information, one may suggest
that the post-perihelion rate of fading of the bright Kreutz
sungrazers depends on the number of major nuclear fragments
seen, $N_{\rm frg}$, satisfying approximately a relation
\begin{displaymath}
n^+ = 4.4 - 0.2 \, N_{\rm frg}. \rlap{\hspace{1.98cm}\mbox{\rm (A-3)}}
\end{displaymath}
If an object does not fragment{\vspace{-0.02cm}} at perihelion, its
\mbox{$N_{\rm frg} = 1$}.  Formula (A-3) does indeed imply \mbox{$n^+$}
of 4.0 for Ikeya-Seki, 3.3 for the 1882 comet, and 4.2 for the 1843
comet.

The preperihelion and post-perihelion absolute magnitudes are
related because the light curve must be continuous at perihelion:
\begin{displaymath}
H_0^- + 2.5 n^- \log q + C = H_0^+ + 2.5 n^+ \log q + C,\;\;\;\;\;\;
 \rlap{\hspace{-0.42cm}\mbox{\rm (A-4)}}
\end{displaymath}
where $q$ is the perihelion distance, \mbox{$C = 5 \log \Delta_q +
\Phi(\psi_q)$} (canceling out), $\Delta_q$ and $\psi_q$ being,
respectively, the geocentric distance and phase angle at perihelion.
With $n^-\!\!\:$ from Equation~(A-2) and $n^+$ from Equation~(A-3),
one finds
\begin{displaymath}
H_0^+ = H_0^- - (1 \!-\! {\textstyle \frac{1}{2}} N_{\rm frg}) \log q.
 \rlap{\hspace{1.26cm}\mbox{\rm (A-5)}}
\end{displaymath}

In order to get an educated guess on the preperihelion absolute
magnitude of Aristotle's comet, I begin with the comets of 1843 and
1882, the most massive third-generation fragments of Aristotle's
comet.  I recently updated their preperihelion absolute magnitudes,
obtaining 3.4--3.5 (Sekanina 2022b).  In another paper{\vspace{-0.03cm}}
(\mbox{Sekanina} 2025) I estimated magnitude $H_0^-$ of the 12th-century
parents of the two giant 19th century sungrazers, or the second-generation
fragments of Aristotle's comet,~at~2.8.  Extrapolation by two
generations up suggests~for~the~pre\-periheliom absolute magnitude of 
Aristotle's comet
\begin{displaymath}
H_0^- = 1.5. \rlap{\hspace{2.7cm}\mbox{\rm (A-6)}}
\end{displaymath}

Next, to assess the value of $n^+$, it is necessary to estimate
the fragmentation status of Aristotle's comet at perihelion.
Disregarding the statement by Ephorus of Cyme about the comet's
breakup into two plamets as meaningless, a chance of nuclear splitting
at perihelion needs to be considered.   In a paper that was testing
the contact-binary model for the Kreutz system (Sekanina \& Kracht
2022), the mode of fragmentation was contemplated that did not involve
any exchange of momentum between the fragments in an exercise as follows.

Consider an object in the proximity of perihelion, moving in a
sungrazing orbit with a velocity $V$.  At a particular point, let
the object split into two fragments without an exchange of momentum.
At the instant of separation they obviously still share the same
orbital velocity, but the heliocentric distances of their centers of
mass now differ by $U_{1,2}$, a finite amount (on the order of a few
kilometers or at most tens of kilometers), a small fraction of the
dimensions of the parent object, whose position of the center of mass
also differs from that of either fragment.  The fragments are bound
to end up in different orbits with uneven orbital periods.  With a
relative precision of 10$^{-8}$, the orbital period of the second
fragment, $P_2$, is related to the orbital period of the first
fragment, $P_1$ (both in years), by
\begin{displaymath}
P_2 = P_1 \!\left(\!1 - \frac{2U_{1,2}}{r_{\rm frg}^2} P_1^{\frac{2}{3}} \!
 \right)^{\!\!-\frac{3}{2}} \!\!\!, \rlap{\hspace{1.46cm}\mbox{\rm (A-7)}}
\end{displaymath}
where $r_{\rm frg}$ is the heliocentric distance at the time of
separation.  Both $U_{1,2}$ and $r_{\rm frg}$ are{\vspace{-0.05cm}}
expressed~in~AU.  Since $r_{\rm frg}$ is unknown, the ratio
\mbox{$2U_{1,2}/r_{\rm frg}^2$} {\vspace{-0.09cm}}can be replaced
for comparison purposes with an expression \mbox{$2\Upsilon_{1,2}/q^2$},
where $\Upsilon_{1,2}$ is the distance along the radius vector between
the centers of mass of the two fragments normalized to an equivalent
case of separation at perihelion.

Recent experimentation with fragments of tidally disrupted Kreutz
sungrazers suggests that their most likely orbital periods fit
the normalized radial distances of
\begin{displaymath}
-5 \; {\rm km} \leq \Upsilon_{1,2} \leq  -3 \;{\rm km}; \,
 3 \; {\rm km} \leq \Upsilon_{1,2} \leq 5 \; {\rm km}.\;\;\;\;\;\;
 \rlap{\hspace{-0.3cm}\mbox{\rm (A-8)}}
\end{displaymath}
When one adopts for the future barycentric~\mbox{orbital}~period
of Aristotle's comet \mbox{$P_1 \!\simeq\! 735$ yr} \mbox{[363.87
$\!\!-\!\!$ ($-$370.95)]} (e.g.,\,Sekanina 2022b,\,Sekanina\,\&\,Kracht
2022)~and,\,from Table~1, \mbox{$q = 0.00680$ AU}, conditions~(A-8)
predict for the orbital periods of potential perihelion-fragmentation
products of Aristotle's comet:\ \mbox{$622 \; {\rm yr} \leq P_2 \leq
663 \; {\rm yr}$} and \mbox{$820 \: {\rm yr} \leq P_2 \leq 887 \:
{\rm yr}$}, respectively.~The suspected~Kreutz sungrazers would have
appeared in the following years:
\begin{displaymath}
\mbox{\rm AD\,251--292; AD\,449--516.}
 \rlap{\hspace{1.5cm}\mbox{\rm (A-9)}}
\end{displaymath}

Over these two periods of time, Hasegawa \& Nakano (2001) include a single
sungrazer candidate in their list, the Chinese comet of AD~252, discovered
on March~23 of that year.  Even though the Chinese-Byzantine comet of
AD~467, discovered on February~6, is not on the list, its motion was
investigated extensively by Mart\'{\i}nez et al.\ (2022) on 
the basis of Chinese (via Ho 1962) as well as European sources; they
assessed rather favorably a chance that this comet was indeed a
sungrazer.

Accordingly, it is reasonable but by no means certain that there existed
two sungrazers, first-generation fragments of Aristotle's comet outside
the Kreutz system, whose \mbox{$|\Upsilon_{1,2}| = 3\!-\!5$ km}.  On the
other hand, one cannot rule out the possibility that, contrary to
expectation, a potential sungrazer at \mbox{$|\Upsilon_{1,2}| > 5$
km}, such as the comet of AD~133 or 607 on the Hasegawa-Nakano list,
may have been a fragment of Aristotle's comet.  In the absence of a
better estimate, I adopt in the following \mbox{$N_{\rm frg} = 3$},
which includes the progenitor of the Kreutz system in AD~363 plus two
non-Kreutz fragments of the Great Comet of 372~BC.

Inserting $N_{\rm frg}$ into Equation~(A-3) I obtain for the rate of
post-perihelion fading
\begin{displaymath}
n^+ = 3.8.  \rlap{\hspace{2.54cm}\mbox{\rm (A-10)}}
\end{displaymath}
I complete the parametrization of the light curve by determining the
post-perihelion absolute magnitude from Equation~(A-5),
\begin{displaymath}
H_0^+ = 0.4. \rlap{\hspace{2.51cm}\mbox{\rm (A-11)}}
\end{displaymath}
The resulting formulas for the adopted visual light curve of
Aristotle's comet are
\begin{displaymath}
H_{\rm app}^- = 1.5 + 10 \log r + 5 \log \Delta + \Phi(\psi)\;\;\;
 \rlap{\hspace{0.22cm}\mbox{\rm (A-12)}}
\end{displaymath}
before perihelion and
\begin{displaymath}
H_{\rm app}^+ = 0.4 + 9.5 \log r + 5 \log \Delta + \Phi(\psi)\;\;\;
 \rlap{\hspace{0.17cm}\mbox{\rm (A-13)}}
\end{displaymath}
after perihelion.  It is possible, if not likely, that these
formulas overestimate the comet's intrinsic brightness at very
small heliocentric distances.  In particular, the magnitude predicted
in Table~13 for the ephemeris~time of January~21.0~UT might be too
bright, even though the forward-scattering effect contributes nearly
4~magnitudes to that brightness.  An alternative
{\vspace{-0.07cm}}to the proposed light curve is to assume that
the $r^{-n^-}\!$ law applies only down to a distance of $r_0$
\mbox{($r_0 > q$)} and that the intrinsic brightness is
constant at distances smaller than $r_0$.  If so, Equation~(A-4)
would have to be modified accordingly and the comet would have
been intrinsically fainter after perihelion.\\[-0.2cm]

%
%
\begin{center}
{\large \bf Appendix B} \\[0.3cm]
LIMITING MAGNITUDE FOR THE NAKED EYE \\[0.3cm]
\end{center}
The conditions for detecting, with the unaided eye, a stellar object
of apparent visual magnitude $H_{\rm app}$, observed from a given site
at a given time, location in the sky, and positions of the Sun and
the Moon, are described by a {\it limiting magnitude\/}, $H_{\rm lim}$,
which is determined by an algorithm developed by Schaefer (1993, 1998).
The limiting magnitude is a function of (i)~the object's solar and
lunar elongations; (ii)~the object's elevation above the local horizon,
as well as the Sun's and Moon's elevations (if applicable); (iii)~the
Moon's phase (if applicable); and (iv)~the atmospheric and other
conditions at the observing site at the time.  The algorithm also
allows for seasonal and long-term effects.

Even though the algorithm applies strictly to stars, the heads of
Kreutz sungrazers are at small heliocentric distances very good
approximations to stellar appearance because they are known to be
small and sharply defined.  Practical application of the method is
greatly facilitated with the use of L.~Bogan's Java Script version
of the computer code at a website shown in the reference to Schaefer
(1998).  In the case of Aristotle's comet~the input is the following
data on the presumed observing site of Athens:\ the geographic
latitude (38$^\circ\!$.00~N), the altitude ($\sim$20 meters); the air
temperature of $-$5$^\circ$C (to account for Aristotle's ``frosty
weather''), the relative humidity ($\sim$70~percent);~and~the~data
under (i) through (iii) above; by clicking on "Calculate
Results"~one~immediately obtains the visual magnitude of the faintest star
visible with the naked eye.  Comparison~of~the~comet's expected
apparent magnitude with the limiting magnitude determines whether
the comet should or should not have been seen under the given
circumstances.{\vspace{0.05cm}}

\begin{center} 
{\footnotesize REFERENCES}
\end{center}
\vspace{0.05cm}
{\footnotesize
\parbox{8.63cm}{Barrett, A.\ A.\ 1978, J.\ Roy.\ Astron.\ Soc.\ Canada,
 72, 81\\[0.03cm]
Cooper, E.\,J.\,1852, Cometic Orbits. Dublin:\,A.\,Thom Publ.,\,196pp\\[0.03cm]
Hasegawa, I.\ 1980, Vistas Astron., 24, 59 \\[0.03cm]
Hasegawa, I.,\,\,\& Nakano, S.\,2001, Publ.\,Astron.\,Soc.\,Japan, 53,
 931\\[0.03cm]
Ho, P.-Y.\ 1962, Vistas Astron., 5, 127 \\[0.03cm]
Katsonopoulou, D., \& Koukouvelas, I.\ 2022, Seism.\ Res.\ Lett.,~93,
{\hspace*{0.25cm}}2401 \\[0.03cm]
Kolia, E.\ 2011, Annu.\ Brit.\ School Athens, 106, 201 \\[0.03cm]
Kronk, G.\ W.\ 1999, Cometography, Volume 1:\ Ancient--1799.
{\hspace*{0.25cm}}Cambridge, UK:\ University Press, 580pp \\[0.03cm]
Marcus, J.\ N.\ 2007, Int.\  Comet Quart., 29, 39 \\[0.03cm]
Mart\'{\i}nez, M.\,J., Marco, F.\,J., Sicoli, P., \& Gorelli, R.\ 2022,~Icarus,
{\hspace*{0.25cm}}384, 115112 \\[0.03cm]
Pingr\'e,\,A.\,G.\,1783,\,Com\'etographie ou Trait\'e historique et
th\'eorique
{\hspace*{0.25cm}}des com\`etes. Tome Premier. Paris:\ L'Imprimerie
Royale \\[0.03cm]
Schaefer, B.\ E.\ 1993, Vistas Astron., 36, 311 \\[0.03cm]
Schaefer, B.\ E.\ 1998, Sky Tel., 95, 57; Java Script code~by~L.~Bogan
{\hspace*{0.25cm}}at {\tt https://www.bogan.ca/astro/optics/vislimit.html}\\[0.03cm]
Seargent, D.\ 2009, The Greatest Comets in History:\ Broom Stars
{\hspace*{0,25cm}}and Celestial Scimitars.  New York:\ Springer Science+Business
{\hspace*{0.25cm}}Media, LLC, 260pp \\[0.03cm]
Sekanina, Z.\ 2002, Astrophys.\ J., 566, 577 \\[0.03cm]
Sekanina, Z.\ 2021, eprint arXiv:2109.01297 \\[0.03cm]
Sekanina, Z.\ 2022a, eprint arXiv:2211.03271 \\[0.03cm]
Sekanina, Z.\ 2022b, eprint arXiv:2202.01164 \\[0.03cm]
Sekanina, Z.\ 2023, eprint arXiv:2310.05320 \\[0.03cm]
Sekanina, Z.\ 2025, eprint arXiv:2505.14662 \\[0.03cm]
Sekanina, Z., \& Kracht, R.\ 2022, eprint arXiv:2206.10827 \\[0.03cm]
Warner, B.\ 1980, Mon.\ Not.\ Astron.\ Soc.\ South Africa, 39, 69 \\[0.03cm]
Webster, E.\ W.\ 2004, Meteorology.{\vspace{-0.075cm}} Translation of
 Aristotle's Mete-\\[0.05cm]
{\hspace*{0.25cm}}orologica, Book 1.6.  Adelaide:\ University of
 Adelaide~eBooks \\
{\vspace{-1cm}}}
\vspace{0.7cm}
\end{document}